\documentclass[aps,prl,showpacs,preprint]{revtex4}

\usepackage{graphicx} 
\usepackage{amsmath}
\usepackage{stmaryrd}
\usepackage{MnSymbol}
\usepackage{setspace}
\usepackage{physics}
\usepackage{mathtools}
\usepackage{amsfonts}
\usepackage{color}
\usepackage{cancel}
\newcommand{\sotimes}{\otimes^{s}}

\begin{document}
\title{Simulation of quantum algorithms using classical probabilistic bits and circuits}
\author{D. D. Yavuz and A. Yadav}
\affiliation{Department of Physics, 1150 University Avenue,
University of Wisconsin at Madison, Madison, WI, 53706}
\date{\today}
\begin{abstract}
We discuss a new approach to simulate quantum algorithms using classical probabilistic bits and circuits. Each qubit (a two-level quantum system) is initially mapped to a vector in an eight dimensional probability space (equivalently, to a classical random variable with eight probabilistic outcomes). The key idea in this mapping is to store both the amplitude and phase information of the complex coefficients that describe the qubit state in the probabilities. Due to the identical tensor product structure of combining multiple quantum systems as well as multiple probability spaces, $n$ qubits are then mapped to a tensor product of $n$ 8-dimensional probabilistic vectors (i.e., the Hilbert space of dimension $2^n$ is mapped to a probability space of dimension $8^n$). After this initial mapping, we show how to implement the analogs of single-qubit and two-qubit gates in the probability space using correlation-inducing operations on these classical random variables. The key defining feature of both the mapping to the probability space and the transformations in this space (i.e., operations on the random variables) is that they are not linear, but instead affine. Using this architecture, the evolution of the $2^n$ complex coefficients of the quantum system can be tracked in the joint fully-correlated probabilities of the polynomial number of random variables. We then give specific procedures for implementing (1) the Deutsch-Jozsa algorithm,  and (2) the Quantum Fourier Transform in the probability space. Identical to the Quantum case, simulating the Quantum Fourier Transform in the probability space requires $O(n)$ probabilistic bits and $O(n^2)$ (i.e., quadratic in the number of quantum bits) operations. 
\end{abstract} 
\maketitle

\section{I. Introduction}

Quantum computing has generated much enthusiasm over the last three decades due to the possibility of solving difficult computational problems more efficiently than any conceivable classical computer \cite{nielsen,divincenzo1,divincenzo2,vazirani,shor,ekert}. One of the main reasons for this enthusiasm is the discovery of Shor's factoring algorithm, which is a polynomial-time algorithm for finding the prime factors of large numbers, of which no efficient classical algorithm is known. A key component of Shor's factoring algorithm is the Quantum Fourier Transform, which achieves the discrete Fourier transform operation on an exponentially large state space with a polynomial number qubits and operations \cite{coppersmith}. It is now understood that, in addition to factoring, quantum algorithms can be used for solving a variety of problems \cite{deutsch}, including efficient data search \cite{grover}, and finding eigenvalues and eigenvectors of large matrices \cite{abrams}. 

Since its inception, there has also been a rigorous debate regarding what constitutes the key ingredient of the computational speed up in quantum algorithms \cite{watrous,wetterich1,wetterich2,duan}. It is clear that exponentially large dimension of the Hilbert space is one of the key ingredients; yet it is also clear that some degree of entanglement and high fidelity of the gates is also essential \cite{waintal,jozsa,nest}. How much entanglement is needed has been the subject of a rigorous debate \cite{jozsa,nest}. To understand the true power of quantum computers, we need to better understand how exactly they differ from their classical counter-parts. Much recent research also indicates that the first truly useful quantum computers will likely use a hybrid approach, where at least some part of the computation is performed classically, using, for example, classical post-processing of quantum measurement outcomes \cite{bauer,monroe,preskillreview}. If at least certain sections of the quantum computation can be replaced with classical algorithms, this may significantly improve the practical applications and the impact of quantum computers. Furthermore, such classical algorithms may be useful in their own right, since they may provide more efficient means of simulating quantum many body systems.

In this paper, we will discuss a new approach for simulating quantum algorithms using classical probabilistic random variables and correlation-inducing operations on these variables (i.e., circuits). The approach builds on our recent work that map quantum systems to classical probabilistic random variables \cite{deniz_rudhy}. In this recent work, we started with the simplest quantum system (a two-level system, i.e., a qubit) and discussed a mapping of the quantum state to a vector in a probability space (Fig.~1). The mapping is one-to-one and preserves all the information encoded in the wavefunction. Not surprisingly, to be able to store all the information encoded in the complex coefficients, we need to increase the dimension of the system: the mapping is to an eight dimensional probabilistic space from the two dimensional Hilbert space (i.e., to a physical classical random variable with eight probabilistic outcomes). 

Once a single-qubit quantum state is mapped, the next key question is whether the evolution of the state can be captured in the probability space. It is well known that an arbitrary evolution of a single qubit wavefunction can be achieved using combinations of Hadamard gates and phase rotations \cite{nielsen}. We showed how these two main operations can be implemented with appropriate transformations of the mapped vector in the probability space (i.e., using appropriate operations on the classical random variable). One key feature of the transformations in the probability space is that they are affine, but not linear.  In our recent manuscript, we also introduced an analogue of the Schrodinger's equation for the wavefunction which lives in a Hilbert space of arbitrary dimension. This is a continuous differential equation that describes the evolution of the vector in the probability space under an effective ``Hamiltonian".

In the current work, we use this recently suggested mapping of quantum systems to probability spaces, and discuss how one can simulate quantum algorithms using classical random variables and correlation-inducing operations on these random variables. As we discussed above, each qubit is initially mapped to a classical random variable with eight probabilistic outcomes (or, to three probabilistic bits, $p$-bits \cite{datta1,datta2}, since three bits are sufficient to produce eight possibilities). Due to the identical tensor product structure of combining multiple quantum systems as well as multiple probability spaces, $n$ qubits are then mapped to a tensor product of $n$ 8-dimensional probabilistic vectors (i.e., the Hilbert space of dimension $2^n$ is mapped to a probability space of dimension $8^n$). After this initial mapping, we show how to implement analogs of single-qubit and two-qubit gates in the probability space using operations on the classical random variables (in other words transformations of the probability state vector). The key defining feature of both the mapping to the probability space and the transformations in this space is that they are not linear, but instead affine.  After this general construction, we give specific procedures for implementing (1) the Deutsch-Jozsa algorithm \cite{deutsch} and (2) the Quantum Fourier transform \cite{coppersmith} in the probability space. Identical to the Quantum case, simulating Quantum Fourier Transform in the probability space using classical random variables requires $O(n^2)$ operations (i.e., quadratic in the number of quantum bits). Remarkably, using this architecture, the evolution of an exponential number of complex coefficients that define the $n$-qubit quantum wavefunction can be tracked in the fully-correlated joint probabilities of the classical random variables. The probabilities contain the information of both the real and the imaginary parts of the complex coefficients. We also show that at the end of the quantum evolution, when a measurement is performed on a Hermitian observable, its' measurement outcomes can be calculated using the same joint probabilities.

The mapping and the simulation that we discuss use classical random variables, and operations on these variables, with a number that scale polynomially with the number of qubits. However, we will not make a statement regarding the true computational efficiency of our simulator. This is because: (1) There may be an exponentially scaling physical resource that is hiding in a certain aspect of our formalism. (2) To evaluate the true computational efficiency of the simulation, a detailed study of noise and error correction is critical. (3) While the measurement outcomes of the quantum system (at the end of the evolution) can be calculated using the joint probability distribution of the classical random variables, it is not clear if this calculation can be performed efficiently under the presence of noise (this is because of the exponentially small probabilities in the joint probability distribution). We will comment on these issues in more detail in the conclusions section below.

Our work has been heavily influenced by the recent investigations of quantum mechanics within the operational framework of probability theories; in particular the pioneering works of Fuchs and colleagues \cite{fuchs1}, Hardy \cite{hardy}, and Barrett \cite{barrett1}. One of the main tools in these investigations is fine-tuned operator classes that allow Symmetric Informationally Complete (SIC) measurements \cite{fuchs2,fuchs3,boyd}. Other related research has tried to place quantum mechanics under the umbrella of probability theories that are more general than classical, sometimes referred to as post-classical theories of probability \cite{hardy,barrett1,barnum,masanes,rau}. This research has identified a rich landscape and the goal is to place quantum mechanics properly in this landscape in order to better understand its unique properties. 

In other related prior work, we note the extensive literature that have attempted to derive some features of quantum mechanics using classical ``toy" theories. A good summary of various toy theories is discussed in, for example, Ref.~\cite{rudolph1}.  Several prominent examples of these are due to Spekkens \cite{spekkens1}, Bell \cite{bell2}, Beltrametti-Bugajski \cite{bugajski}, Kochen-Specker \cite{kochen}, Aaronson \cite{aaronson}, and Aerts \cite{aerts}. Of particular importance to this work is Aaronson's model \cite{aaronson}, which discusses representing the quantum state as a vector of probabilities, and mapping this vector to another set of probabilities using an appropriate matrix. However, when only represented as a vector of projected probabilities, such a matrix inevitably depends on the initial state of the wavefunction, which is very different from the approach that we consider here. 

This work is also related to the mapping of quantum states to probability-like distributions, typically referred to as quasiprobabilities \cite{bartlett1,bartlett2,bartlett3,raussendorf1,raussendorf2,eisert,zhu,wootters}. The most well-known example of a quasiprobability distribution is the Wigner function. It is well-known that quasi-probabilities can have negative values; in fact, the true quantum mechanical nature of the wavefunction is expressed in these negative regions. We argue that when one allows for maps and transformations that are not necessarily linear, one can capture a quantum state (as well as its' evolution) using only probabilities (i.e., negative values are not needed). We commented on these connections more thoroughly in our recent manuscript \cite{deniz_rudhy}. In the current paper, we will focus specifically on simulating quantum algorithms using this approach, such as the Deutsch-Jozsa algorithm and the Quantum Fourier Transform.

\section{II. Preliminaries}
In traditional formulation of quantum mechanics, a quantum state is described by a complex wavefunction, $|\psi \rangle$, in a Hilbert space, $H$. This state will evolve according to Schrodinger's equation, which conserves the norm of the wavefunction. This time evolution of the quantum state can be described using an appropriate unitary matrix, $\hat{U}$, that satisfies, $\hat{U}^\dagger \hat{U} = \hat{U} \hat{U}^\dagger = \hat{I}$. With this evolution, the state is mapped to $|\psi \rangle \longrightarrow \hat{U} | \psi \rangle$. 

In any classical probabilistic experiment, we will have a set of probabilities, which we can also think of as constituting a vector, in a probabilistic space, $S$. We will denote such a probabilistic vector with $\vec{s}$. Each of the entries of this vector has to be between 0 and 1, i.e., $0 < s_i <1$, and furthermore, the entries need to sum to unity, $\sum_i s_i =1$. Because the entries add up to unity, such a vector lies on certain surface in the probabilistic space, and this surface is called the simplex \cite{definetti}. Similar to a quantum state, such a probabilistic vector can also evolve in time (for example, because of a change in  the experimental conditions). We can view such evolution as mapping a vector in space $S$, to another vector. We will denote such mapping with $ T : S \longrightarrow S$. Usually, such evolution is described by multiplying the vector $\vec{s}$ with a Stochastic matrix, $\tilde{\mathcal{M}}$, i.e., $T (\vec{s}) = \tilde{\mathcal{M}} \cdot \vec{s}$. A stochastic matrix is a matrix whose columns sum up to 1. This assures that the resultant vector also is normalized; i.e., its' components add up to unity.  Throughout this paper, all quantum mechanical operators will be presented by a hat (for example $\hat{U}$), whereas all the  transformations of the simplex vectors will be presented by a tilde (for example, $\tilde{\mathcal{M}}$).

The probabilistic vectors can also undergo affine transformations of the form $T (\vec{s}) = \vec{a} + \tilde{M} \cdot \vec{s} $. Here, the constant ``offset" vector $\vec{a}$ and the matrix $\tilde{M}$ should be chosen such that the mapped vector $ T ( \vec{s} ) $ is a valid probability distribution. The conditions for the matrix $\tilde{M}$ such that $ T : S \longrightarrow S$ is a valid transformation is different from stochasticity. We will discuss these conditions in detail below. Affine nature of the transformations require that a statistical mixture of the input vectors should produce the same statistical average of the transformed vectors. More formally, for any two vectors $\vec{s}$ and $\vec{s}'$, and two constants $\lambda$ and $\lambda'$ such that $\lambda + \lambda' =1 $, we have $T (\lambda \vec{s} + \lambda' \vec{s}' ) = \lambda T ( \vec{s} ) + \lambda' T (\vec{s}')$. This type of affine transformations of probabilistic vectors has not received much attention before, and is one of the central ideas of this work.

Figure~1 summarizes the key features of our approach. We start with mapping a single qubit wavefunction to the probability simplex. Not surprisingly, to be able to store all the information that is encoded in the complex coefficients, we need to increase the dimension of the system. As shown in Fig.~1(a), the mapping is to an eight dimensional probability space. In this mapping of the qubit, we have something quite physical in mind: that is, the mapping is to a physical classical random variable with eight probabilistic outcomes. This can, for example, be visualised as a ``die" with eight faces and the vector $\vec{s}$ stores the probabilities in these eight outcomes. The key is that these 8 probabilities store both the amplitude and the phase information in the complex qubit wavefunction. When a measurement is made on the quantum wavefunction, one finds the quantum system to be in one of the two states with probabilities given by the magnitude square of the complex coefficients. Because the map stores both the phases and the amplitudes of these complex coefficients, not surprisingly, by measuring these probabilities (i.e., by repeatedly throwing the die and measuring the components of the 8-dimensional vector $\vec{s}$), one can also uniquely calculate the probabilistic outcomes of the quantum system.

\begin{figure}[h]
\vspace{-0cm}
\begin{center}
\includegraphics[width=0.95\textwidth]{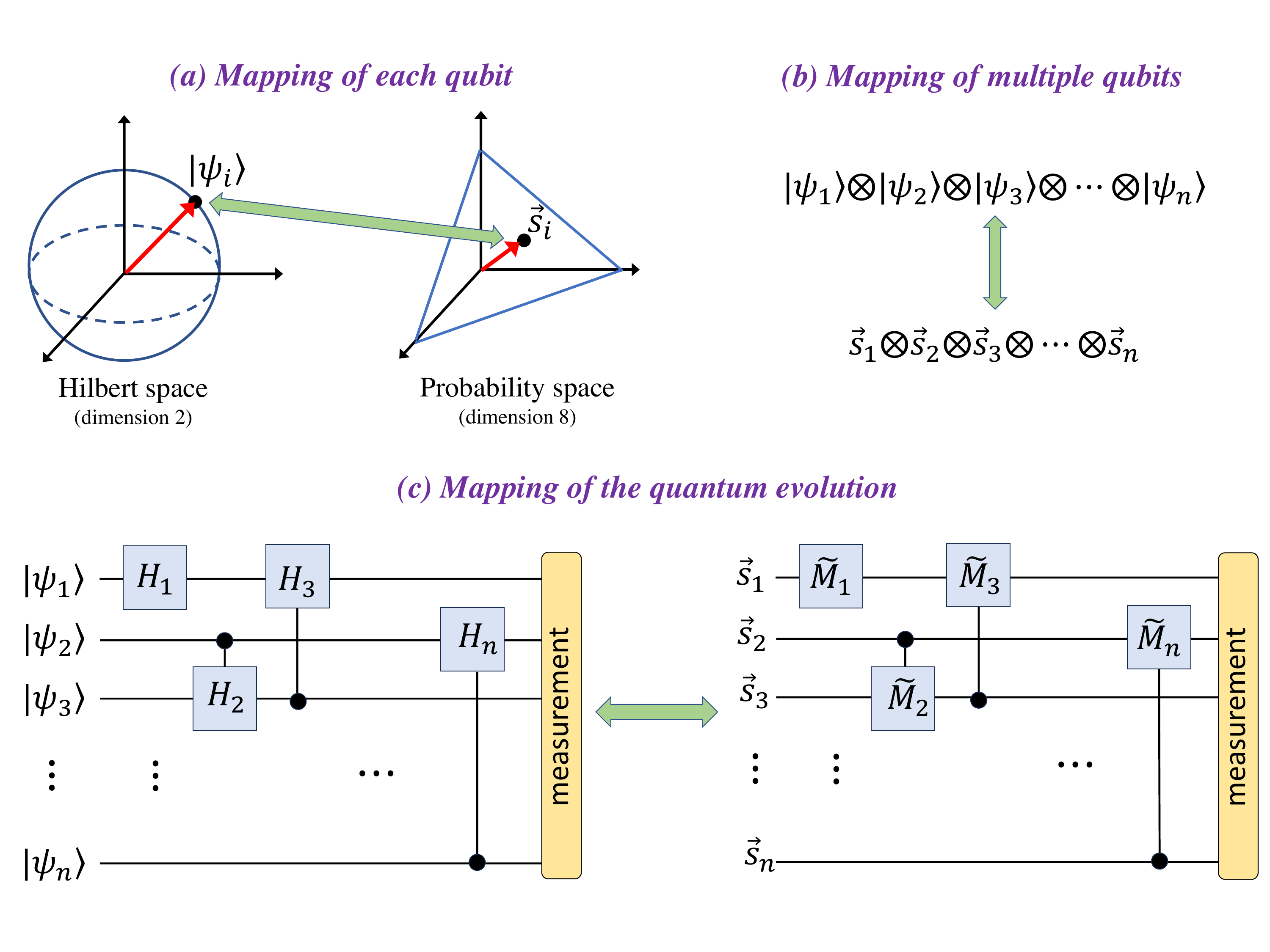}
\vspace{-0.8cm} 
\caption{\label{scheme} \small The simplified schematic of the approach that we will study in this work. (a) We start with a qubit with wavefunction, $| \psi_j \rangle $, and discuss a one-to-one mapping, $\varphi$ from the Hilbert space $H$ to a vector $\vec{s}_j$ in an eight dimensional probability space $S$ (which is a real Euclidean space). In this mapping of the qubit, we have something quite physical in mind: that is, the mapping is to physical classical random variable with 8 values. This can, for example, be visualised as a ``die" with 8 faces; the vector $\vec{s}$ stores the probabilities in these 8 outcomes. (b) The wavefunction for a multi-partite quantum system is initially mapped to a tensor product of individual simplex vectors. This is due to the identical tensor product rule of combining multiple quantum systems as well as multiple probability spaces. (c) With the initial product wavefunction (unentangled state), a quantum algorithm runs through a sequence of single-qubit and two-qubit gates. Each of these operations can be mapped to a corresponding affine transformation in the probability space. The end result is that, the quantum evolution in the Hilbert space of dimension $2^n$ (i.e., an exponentially large number of complex coefficients), can be smoothly tracked in the probability space of dimension $8^n$. At the end of the evolution, the fully-correlated joint probabilities in the $\vec{s}$ vectors are measured. Because these probabilities contain both the amplitude and the phase information in the $2^n$ complex coefficients of the quantum system, not surprisingly, the measurement outcomes of the quantum system can be calculated by measuring these joint probabilities. }
\end{center}
\vspace{-0.3cm}
\end{figure}

Throughout this manuscript, we will formulate our approach using these above-mentioned eight-dimensional probabilistic vectors. However, we note that, our whole scheme can instead be formulated in terms of classical random variables with only two outcomes such as a coin-flip (i.e., a probabilistic bit, or a $p$-bit.) Eight possible outcomes require three $p$-bits and one can visualize the mapping of the qubit to three physical $p$-bits (instead of mapping to a single ``die" with eight outcomes).

We will then consider multiple qubit systems. Here, each qubit wavefunction, $\ket{\psi_j}$ is mapped to an eight-dimensional simplex vector, $\vec{s}_j$. Due to the identical tensor product structure of combining multi-partite systems, the combined wavefunction of the initial multi-qubit system is mapped to a tensor product of simplex vectors (this can be understood intuitively as the joint probabilities of several events happening together). This is schematically shown in Fig.~1(b). In this multi-partite mapping, we again have something quite physical in mind. $n$ qubits are mapped into $n$ 8-dimensional dice (or equivalently, to $3n$ probabilistic bits), and initially, the information in the quantum wavefunction is stored in the tensor product of probabilities stored in the corresponding $\vec{s}$ vectors.

With this initial mapping, the next question that we address is if the evolution of the quantum system can be captured in the probability simplex. For this purpose, we will first discuss how to implement analogs of single-qubit and two-qubit gates in the probability simplex. Each gate in the quantum system can be viewed as changing the values of the complex coefficients in the Hilbert space. What is remarkable is that, these modifications of the complex coefficients (due to the quantum evolution) can be fully captured using corresponding affine transformations acting on single-simplex or two-simplex vectors. These affine operations can be viewed as physical experimental operations, that change the probabilistic outcomes of a single ``die", or specific operations that induce correlations between the two ``dice".

With this construction, we will then shift our focus to implementing specific quantum algorithms. Here, since any quantum algorithm can be implemented using a sequence of single-qubit and two-qubit gates, we basically track the algorithm in the probability space using a sequence of affine transformations. The end result is that, the quantum evolution in the Hilbert space of dimension $2^n$ (i.e., an exponentially large number of complex coefficients), can be smoothly tracked in the probability space of dimension $8^n$. This is schematically shown in Fig.~1(c). We will specifically focus on the Deutsch-Josza algorithm and the Quantum Fourier Transform (which is the foundation of Shor's factoring algorithm).

\section{III. Mapping of Single Qubit Wavefunction and its' evolution}
We first discuss mapping of the single-qubit wavefunction and its evolution in the Hilbert space. This section will follow closely the discussion in our recent manuscript \cite{deniz_rudhy}, which we include here for completeness. In the following subsections, we describe the mapping of the wavefunction from the Hilbert space to the probability space, $\varphi \ket{\psi}$ and also mapping of the wavefunction evolution under unitary operator $\hat{U}$, $\tilde{M} [\hat{U}]$.

\subsection{Mapping of the single-qubit wavefunction}
For a single qubit, we can decribe the state $\ket{\psi}$ in the logical qubit basis as,
\begin{equation}
    \ket{\psi}=c_0\ket{0}+c_1\ket{1}\equiv
    \left( \begin{array}{c} 
    c_0 \\
    c_1 
    \end{array} \right) \equiv 
    \left( \begin{array}{c} 
    x_0 + i y_0 \\
    x_1 + i y_1
    \end{array} \right)=\vec{x}+i\vec{y} \quad . 
\end{equation}
Here, the states $| 0 \rangle $ and $| 1 \rangle$ are the logical states, and  $c_0$ and $c_1$ are the complex coefficients satisfying the usual normalization condition, $| c_0|^2 + |c_1|^2 =1$. In what follows, instead of the complex coefficients $c_0,c_1$, we will work with their real and imaginary parts $\vec{x},\vec{y}$, which are two-dimensional vectors defined as:
\begin{eqnarray}
\vec{x} \equiv \Re\ket{\psi} = \left( \begin{array}{c} x_0  \\ x_1 \end{array} \right)  \quad ,  \quad 
\vec{y} \equiv \Im\ket{\psi}  = \left( \begin{array}{c} y_0 \\ y_1 \end{array} \right)  \quad .
\end{eqnarray}
We propose the following mapping $\varphi:H\mapsto S$ of the quantum state $|\psi \rangle$ in Hilbert space $H$ to a vector $\vec{s}$ in the probability space $S$:
\begin{eqnarray}
\varphi\ket{\psi}=\varphi(\vec{x}+i\vec{y})=\vec{s}=\frac{1}{8}\left(\begin{array}{c}
	1\\
	1\\
	\vdots\\
	1
\end{array}\right)+\frac{1}{8}\left(\begin{array}{r}
	\vec{x}\\
	-\vec{x}\\
	\vec{y}\\
	-\vec{y}
\end{array}\right) \equiv \frac{1}{8}(\vec{u}+\vec{p}) \quad .  
\end{eqnarray}

Here, we have defined a vector with uniform entries $\vec{u} \equiv \vec{1}$ and also another vector that stores the deviation of the probabilities from the uniform distribution, $\vec{p} \equiv 8 \vec{s} - \vec{1}$.
We note that the vector $ \vec{s}$, as defined above, represents a valid probability distribution. That is, each of the entries is between 0 and 1 (i.e., $ 0 < s_i <1$), and these entries sum up to unity, $\sum_i s_i =1$. The fact that we need to increase the dimension from 2 to 8 is intuitive. For each complex coefficient, we need to store two real numbers, the real part and the imaginary part. Furthermore, for each real number, we need to store the quantity with both signs. This is because, in order to map the transformations of the quantum state, we will need access to both signs of these coefficients. Hence, the factor of 4 increase in the dimension. The map is injective (i.e., one-to-one), but not surjective. The main insight in the mapping of Eq.~(3) is that the phase and the amplitude information (for the real and imaginary parts of the complex coefficients) can be stored in how much the probabilities deviate from purely random quantity (hence the initial ``1" in all the entries of $\vec{s}$). 

A key property of the mapping of Eq.~(3) is that it is not linear. By inspection, a superposition of two wavefunctions do not map to the same superposition of their mapped vectors: $\varphi(a\ket{\psi}+b\ket{\phi})\neq a\varphi\ket{\psi}+b\varphi\ket{\phi}$ for $\ket{\psi},\ket{\phi}\in H; a,b\in \mathbb{C}$. A more explicit expression for the map, which clearly shows its affine (but not linear) nature, is:
\begin{align}
    &\varphi\ket{\psi}=\frac{1}{8}(\vec{u}+\vec{\gamma}\otimes\Re\ket{\psi}+\vec{\gamma}'\otimes\Im\ket{\psi})\\
    &\vec{\gamma}=\left(\begin{array}{r}
	1\\
	-1\\
	0 \\
	 0  \end{array}  \right) \quad , \quad 
  \vec{\gamma}'=\left(\begin{array}{r}
	0 \\
	0 \\
	1 \\
	 -1  \end{array}  \right) \quad . 
\end{align}

We note that the set of states $\vec{s}$ defined in the simplex by Eq.~(3) form a convex surface. That is, for two different states $\vec{s}$ and $\vec{s}'$, and for coefficients $\lambda$ and $\lambda'$ such that $\lambda + \lambda' =1$, any combination $ \lambda \vec{s} + \lambda' \vec{s}'$ is also an allowed mapped state. This is similar to what is discussed in Refs.~\cite{barrett1,hardy}. We also note, however, that, differing from the prior work, the simplex vector with all of its' entries equal to 0 (which we can denote by $\vec{0}$) is not a valid mapped vector. Even if we were to include not-normalized quantum states (where the probabilities leek out of the system, for example), in the limit, $x_i \rightarrow 0, y_i \rightarrow 0$, all of the entries for the vector in the simplex would approach $\frac{1}{8}$, i.e., $ \vec{s} \rightarrow \frac{1}{8} \vec{u}$.

When we discuss analogs of two-qubit gates, as well as quantum algorithms below, it will be useful to use a notation analogous to the logical $| 0 \rangle$ and $|1 \rangle$ quantum states. For this purpose, we introduce the simplex vectors, $\vec{s}_0$ and $\vec{s}_1$, and correspondingly, $\vec{p}_0$ and $\vec{p}_1$. These are the vectors that are obtained by mapping the quantum state $\ket{\psi} = \ket{0}$, and $\ket{\psi} = \ket{1}$, respectively. More explicitly, these vectors are:
\begin{eqnarray}
       \varphi\ket{0} & = & \vec{s}_0 \equiv \frac{1}{8} (\vec{u}+\vec{p}_{0}),\quad \varphi\ket{1}=\vec{s}_1 \equiv \frac{1}{8} (\vec{u}+\vec{p}_{1}) \quad , \nonumber \\
       \vec{p}_0 & \equiv & \left(\begin{array}{r}
	1\\ 	
    0\\
	-1 \\
	 0  \\
0 \\
0 \\ 
0 \\
0
  \end{array}  \right) ,  \quad 
        \vec{p}_1 \equiv \left(\begin{array}{r}
	0\\ 	
    1\\
	0 \\
	-1  \\
0 \\
0 \\ 
0 \\
0
  \end{array}  \right) \quad .   
\end{eqnarray}
The above mapped vectors from the logical $\ket{0}$ and $\ket{1}$ are sufficient when the quantum algorithm only requires real coefficients in the quantum state (such as the Deutsch-Jozsa algorithm). However, when imaginary components of the coefficients are necessary, the above vectors are not sufficient. We, therefore, introduce a more general version of these vectors, $\vec{P}_0$ and $\vec{P}_1$, which will be critical in the discussion of the Quantum Fourier Transform. Unlike the constant vectors $\vec{p}_0$ and $\vec{p}_1$, we allow these more general vectors to be a function of a complex number, $c$. For any complex coefficient, $c$, these two vectors are defined as:
\begin{eqnarray}
       \vec{P}_0 (c)  \equiv  \left(\begin{array}{c}
	\Re(c)\\ 	
    0\\
	-\Re(c) \\
	 0  \\
\Im(c) \\
0 \\ 
-\Im(c) \\
0
  \end{array}  \right) ,  \quad 
       \vec{P}_1 (c)  \equiv  \left(\begin{array}{c}
	0 \\ 	
    \Re(c)\\
	0 \\
	 -\Re(c)  \\
0 \\
\Im(c) \\ 
0  \\
-\Im(c)
  \end{array}  \right)  \quad . 
\end{eqnarray}
which can be abstractly expressed in one statement as, 
\begin{equation}
    \vec{P}_{b}(c)=[\Re(c)\vec{\gamma}+\Im(c)\vec{\gamma}']\otimes\ket{b} \quad .
\end{equation}
for a given logical qubit state $\ket{b}, b\in\mathbb{B}=\{0,1\}$. These vectors allow us to express the map for a more general single qubit state, $\ket{\psi}=c_0\ket{0}+c_1\ket{1}$, in the following simplified form: 
\begin{equation}
    \varphi\ket{\psi}=\vec{s}(\psi)\equiv\frac{1}{8}[\vec{u}+\vec{P}_{0}(c_0)+\vec{P}_{1}(c_1)]=\frac{1}{8}[\vec{u}+\sum_{b\in\mathbb{B}}\vec{P}_{b}(c_b)] \quad . 
\end{equation}
We note that $\vec{P}_{0}(1)=\vec{p}_{0}$ and $\vec{P}_{1}(1)=\vec{p}_{1}$,   whereas $\vec{P}_{0}(0)= \vec{P}_1(0) = \vec{0}$. We also have:  
\begin{align}
    &\vec{P}_{0}(r e^{i\phi})=r\vec{P}_{0}(e^{i\phi}), \quad
    \vec{P}_{1}(r e^{i\phi})=r\vec{P}_{1}(e^{i\phi}) \quad , \\
    \text{and, }&\vec{P}_0(r)=r\,\vec{p}_{0}, \quad \vec{P}_1(r)=r\,\vec{p}_{1},  \quad \forall r\in\mathbb{R}\quad .
\end{align}
We finally note that the map $\vec{P}_b:\mathbb{C}\mapsto \mathbb{R}^8$ also satisfies the following additive property:
\begin{equation}
    \vec{P}_{b}(\sum_{k}c_k)=\sum_{k}\vec{P}_{b}(c_k), \forall c_k\in \mathbb{C} \quad . 
\end{equation}

\subsection{Single qubit transformations in the simplex}
The central question is what type of transformations of the probability vector, $T : S \rightarrow S$, should we be looking for. Motivated by the mapping of Eq.~(3), we look for affine transformations of the simplex vector of the form a translation added on linear combinations of the simplex vector entries. Note that the entries of $\vec{p}$ in Eq.~(3) sum up to zero; i.e., $\sum_i p_i=\vec{u}\cdot\vec{p}=0$. Furthermore, the Euclidian norm of $\vec{p}$ is a constant $||\vec{p}||=\sqrt{2}$, since we have $x_0^2+y_0^2+x_1^2+y_1^2 =1$ (this is because of the normalization of the state $\ket{\psi}$). We also note that the two vectors that form the simplex vector $\vec{s}$ are orthogonal to each other, $\vec{u} \cdot \vec{p} =0$. As a result, we have $ ||\vec{s}||=\sqrt{||\vec{u}||^2+||\vec{p}||^2}/8=\sqrt{10}/8$, which is also constant. This shows that $\vec{s}$ lies on the intersection of a seven-dimensional hypersphere, with four seven-dimensional hyperplanes, resulting in a three dimensional hypersurface $S$. 

As it will be clear below, because the quantum gates form linear combinations of the entries of $\vec{p}$, we first view the mapping of the simplex vector $\vec{s}$, as instead mapping $\vec{p}$ to another vector. We will call the matrix for this mapping to be $\tilde{M}[\hat{U}]$ (corresponding to the unitary quantum evolution $\hat{U}$):
\begin{eqnarray}
\ket{\psi}\longrightarrow \hat{U}\ket{\psi}\iff\vec{p} \longrightarrow \tilde{M}[\hat{U}] \cdot \vec{p} \quad . 
\end{eqnarray}
Expressed as $T[\hat{U}]$ acting on the full simplex state $\vec{s}$, the transformation of Eq.~(13) is, $T[\hat{U}](\vec{s}) = \frac{1}{8} \left( \vec{u} + \tilde{M}[\hat{U}] \cdot \vec{p} \right)$, which gives $ T[\hat{U}](\vec{s}) = \frac{1}{8} \left[ \vec{u} + \tilde{M}[\hat{U}] \cdot \left( 8 \vec{s} -  \vec{u} \right) \right] $, or writing it slightly differently, 
\begin{eqnarray}
T[\hat{U}](\vec{s}) =\frac{1}{8} \left(\tilde{I}_{8\times8} - \tilde{M}[\hat{U}] \right) \cdot \vec{u} + \tilde{M}[\hat{U}] \cdot \vec{s} \quad .  
\end{eqnarray}
\noindent Here, the quantity $\tilde{I}_{8\times8}$ is the $ 8 \times 8$ identity matrix. Below, we will give explicit general expressions for the $8 \times 8$ matrices, $\tilde{M}[\hat{U}]$, tracking a specific evolution, $\hat{U}$, of the quantum state. With the matrix $\tilde{M}$ given, Eq.~(14) describes the explicit transformation of the probability vector, with the map $T : S \longrightarrow S$ in the simplex. 

We note that, the first term in the right hand side of Eq.~(14) is a translation for each of the entries of the vector (an offset). Because of this term, the map $T : S \longrightarrow S$ is not linear (i.e., the sum of two vectors $\vec{s}$ and $\vec{s}'$ would not transform as the sum of the individual transforms). However, $T$ is an affine map. For two vectors, $\vec{s}$ and $\vec{s}'$, and for coefficients $\lambda$ and $\lambda'$ such that $\lambda + \lambda' =1$, we have $ T( \lambda \vec{s} + \lambda' \vec{s}') = \lambda T(\vec{s}) + \lambda' T(\vec{s}') $. 

The constraints on the matrix $\tilde{M}$ of above such that $T : S \longrightarrow S$  is a valid map is different from stochasticity. Specifically, the two necessary constraints are (1) $\tilde{M}$ should be such that the norm of the resulting vector is preserved since we need to have: $ || \tilde{M} \cdot \vec{p} || = \sqrt{2}$. Because of the specific form for the vector $\vec{p}$, this norm conservation does not imply orthogonality of the matrix $\tilde{M}$. By inspection, the necessary constraint is that the sum of the squares of the entries in each row must add up to unity: i.e., $\sum_j \tilde{M}_{{ij}}^2 =1$ for each row $i$.  (2) The rows of $\tilde{M}$ should be related to each other such that the entries of $ \tilde{M} \cdot \vec{p}$ sum up to zero. Specifically, $ \tilde{M} \cdot \vec{p}$ should produce a column vector of the form shown in Eq.~(3), with respective entries having equal amplitude and opposite signs. This assures that the resulting full simplex vector, $\frac{1}{8} \left( \vec{u} +\tilde{M} \cdot \vec{p} \right)$ is  a valid probability distribution (i.e., its' entries add up to unity). 

Given a general unitary matrix $\hat{U}$ acting on a quantum state vector $\ket{\psi}$, we note that the real and imaginary parts of the wavefunction will transform as: 
\begin{equation}
	\hat{U}\ket{\psi}=\left[\Re(\hat{U})+i\Im(\hat{U})\right]\cdot(\vec{x}+i\vec{y})=\left[\Re(\hat{U})\cdot\vec{x}-\Im(\hat{U})\cdot\vec{y}\right]+i\left[ \Re(\hat{U})\cdot\vec{y}+\Im(\hat{U})\cdot\vec{x}\right] \quad . 
\end{equation}
\noindent Here, the quantities $\Re(\hat{U})$ and $\Im(\hat{U})$ are the real and imaginary components of the evolution operator $\hat{U}$, respectively. This implies that, under general unitary evolution, the real and imaginary parts of the wavefunction will evolve as:
\begin{eqnarray}
	\vec{x} & \longrightarrow & \left[\Re(\hat{U})\cdot\vec{x}-\Im(\hat{U})\cdot\vec{y} \right] \quad \nonumber \\
	\vec{y} & \longrightarrow & \left[\Re(\hat{U})\cdot\vec{y}+\Im(\hat{U})\cdot\vec{x} \right] \quad . 
\end{eqnarray}
For the mapped vector $\vec{s}$ in the simplex, the above evolution of the real and imaginary parts of the wavefunction implies the following transformation of the vector $\vec{p}$:
\begin{equation}
	\vec{p}=\left(\begin{array}{r}
		\vec{x}\\
		-\vec{x}\\
		\vec{y}\\
		-\vec{y}
	\end{array}\right)\longrightarrow \left(\begin{array}{c| c| c| c}
		\Re(\hat{U})&O&O&\Im(\hat{U})\\
        \hline
		O&\Re(\hat{U})&\Im(\hat{U})&O\\
        \hline
		\Im(\hat{U})&O&\Re(\hat{U})&O\\
        \hline
		O&\Im(\hat{U})&O&\Re(\hat{U})
	\end{array}\right)\left(\begin{array}{r}
		\vec{x}\\
		-\vec{x}\\
		\vec{y}\\
		-\vec{y}
	\end{array}\right)=\tilde{M}(\hat{U})\cdot\vec{p} \quad . 
\end{equation}
We also note that due to the structure of $\vec{p}$, the following two transformations are equivalent:
\begin{equation}
    \left(\begin{array}{c| c| c| c}
		\Re(\hat{U})&O&O&\Im(\hat{U})\\
        \hline
		O&\Re(\hat{U})&\Im(\hat{U})&O\\
        \hline
		\Im(\hat{U})&O&\Re(\hat{U})&O\\
        \hline
		O&\Im(\hat{U})&O&\Re(\hat{U})
	\end{array}\right)\left(\begin{array}{r}
		\vec{x}\\
		-\vec{x}\\
		\vec{y}\\
		-\vec{y}
	\end{array}\right)=\left(\begin{array}{c| c| c| c}
		O&-\Re(\hat{U})&O&\Im(\hat{U})\\
        \hline
		-\Re(\hat{U})&O&\Im(\hat{U})&O\\
        \hline
		\Im(\hat{U})&O&O&-\Re(\hat{U})\\
        \hline
		O&\Im(\hat{U})&-\Re(\hat{U})&O
	\end{array}\right)\left(\begin{array}{r}
		\vec{x}\\
		-\vec{x}\\
		\vec{y}\\
		-\vec{y}
	\end{array}\right) \quad . 
\end{equation}
As a result of the above equivalence, we consider both of the matrices to be equivalent definitions of the transformation $\tilde{M}[\hat{U}]$, associated to a general evolution of the state:
\begin{equation}
    \label{m1}
	\tilde{M}[\hat{U}]=\left(\begin{array}{c| c| c| c}
		\Re(\hat{U})&O&O&\Im(\hat{U})\\
        \hline
		O&\Re(\hat{U})&\Im(\hat{U})&O\\
        \hline
		\Im(\hat{U})&O&\Re(\hat{U})&O\\
        \hline
		O&\Im(\hat{U})&O&\Re(\hat{U})
	\end{array}\right)=\tilde{I}_{4\times4}\otimes\Re(\hat{U}) + \tilde{\Lambda}\otimes\Im(\hat{U}) \quad ,
\end{equation}
or, 
\begin{equation}
    \label{m2}
	\tilde{M}[\hat{U}]=\left(\begin{array}{c| c| c| c}
		O&-\Re(\hat{U})&O&\Im(\hat{U})\\
        \hline
		-\Re(\hat{U})&O&\Im(\hat{U})&O\\
        \hline
		\Im(\hat{U})&O&O&-\Re(\hat{U})\\
        \hline
		O&\Im(\hat{U})&-\Re(\hat{U})&O
	\end{array}\right)=-\tilde{\Lambda}^2\otimes\Re(\hat{U}) + \tilde{\Lambda}\otimes\Im(\hat{U}) \quad .
\end{equation}
where,
\begin{equation}
    \tilde{\Lambda}=\left(\begin{array}{c c c c}
		0&0&0&1\\
		0&0&1&0\\
		1&0&0&0\\
		0&1&0&0
	\end{array}\right), \quad \tilde{\Lambda}^2=\left(\begin{array}{c c c c}
		0&1&0&0\\
		1&0&0&0\\
		0&0&0&1\\
		0&0&1&0
	\end{array}\right)
\end{equation}
Using the equivalence between Eq.~(\ref{m1}) and Eq.~(\ref{m2}) we can prove the useful property that $\tilde{M}[\hat{U}_1\hat{U}_2]$ and $\tilde{M}[\hat{U}_1]\tilde{M}[\hat{U}_2]$ have the same action over a particular $\vec{p}$ state, described in what follows,
\begin{align}
    &\nonumber \tilde{M}[\hat{U}_1]\tilde{M}[\hat{U}_2]=(\tilde{I}_{4\times4}\otimes\Re(\hat{U}_1) + \tilde{\Lambda}\otimes\Im(\hat{U}_1))(\tilde{I}_{4\times4}\otimes\Re(\hat{U}_2) + \tilde{\Lambda}\otimes\Im(\hat{U}_2))\\
    &\nonumber =\tilde{I}_{4\times4}\otimes\Re(\hat{U}_1)\Re(\hat{U}_2)+\tilde{\Lambda}\otimes(\Re(\hat{U}_1)\Im(\hat{U}_2)+\Im(\hat{U}_1)\Re(\hat{U}_2))+\tilde{\Lambda}^2\otimes(\Im(\hat{U}_1)\Im(\hat{U}_2))\\
    &\nonumber \equiv \tilde{I}_{4\times4}\otimes(\Re(\hat{U}_1)\Re(\hat{U}_2)-\Im(\hat{U}_1)\Im(\hat{U}_2))+\tilde{\Lambda}\otimes(\Re(\hat{U}_1)\Im(\hat{U}_2)+\Im(\hat{U}_1)\Re(\hat{U}_2))\\
    &\equiv \tilde{I}_{4\times4}\otimes\Re(\hat{U}_1\hat{U}_2)+\tilde{\Lambda}\otimes\Im(\hat{U}_1\hat{U}_2)=\tilde{M}[\hat{U}_1\hat{U}_2]
\end{align}
This identity implies that, 
\begin{equation}
    T[\hat{U}_1]\circ T[\hat{U}_2]=T[\hat{U}_1\hat{U}_2]
\end{equation}
which is in direct analogy to compositions of two or more unitary operations on any qubit state $\ket{\psi}$. Moreover, this in turn shows that the transforms $T[\hat{U}]$ for any given unitary $\hat{U}$ are reversible as they should be for closed quantum systems: $T[\hat{U}]\circ T[\hat{U}^{\dagger}]=T[\hat{U}^{\dagger}]\circ T[\hat{U}]=T[\hat{I}_{2\times2}]=\text{identity}$ (i.e., $T[\hat{U}^{\dagger}]=T^{-1}[\hat{U}]$). 

Furthermore, restricted to the simplex manifold $S$ the transform $T[\hat{U}]$ for any unitary $\hat{U}$ is affine, i.e., 
\begin{equation}
    T[\hat{U}](\lambda\vec{s}+(1-\lambda)\vec{s}')=\lambda\,T[\hat{U}](\vec{s})+(1-\lambda)\,T[\hat{U}](\vec{s}'),\quad 0\leqslant\lambda\leqslant 1
\end{equation}
As a specific example, we next discuss how to implement the analog of Hadamard gate on a single qubit. A Hadamard gate is accomplished by multiplying the state vector $| \psi \rangle$ with the following unitary matrix \cite{nielsen}:
\begin{eqnarray}
\hat{H} = \left(\begin{array}{rr} 
\frac{1}{\sqrt{2}} & \frac{1}{\sqrt{2}} \\
\frac{1}{\sqrt{2}} & -\frac{1}{\sqrt{2}} 
\end{array} \right) \quad .
\end{eqnarray}
The effect of the Hadamard gate on the quantum state, explicitly expressed in terms of the real and imaginary parts of the complex coefficients, is:
\begin{eqnarray}
| \psi \rangle & \longrightarrow & \hat{H} | \psi \rangle \nonumber \\
\left( \begin{array}{c} 
x_0 + i y_0 \\
x_1 + i y_1
\end{array} \right) & \longrightarrow &  
\left( \begin{array}{c} 
\frac{1}{\sqrt{2}} x_0 +  \frac{1}{\sqrt{2}} x_1 + i \left( \frac{1}{\sqrt{2}} y_0 +  \frac{1}{\sqrt{2}} y_1 \right) \\
\frac{1}{\sqrt{2}} x_0 -  \frac{1}{\sqrt{2}} x_1 + i \left( \frac{1}{\sqrt{2}} y_0 -  \frac{1}{\sqrt{2}} y_1 \right) 
\end{array} \right)  \quad . 
\end{eqnarray}
By inspection, the required $ 8 \times 8$ matrix for the transformation of Eq.~(26) is:
\begin{eqnarray}
\tilde{M}[\hat{H}] = \left( \begin{array}{cc|cc|cc|cc}
\frac{1}{\sqrt{2}} & \frac{1}{\sqrt{2}} & 0 & 0 &  0 & 0 & 0 & 0 \\
\frac{1}{\sqrt{2}} & -\frac{1}{\sqrt{2}} & 0 & 0 & 0 & 0 & 0 & 0 \\ 
\hline
0 & 0 & \frac{1}{\sqrt{2}} & \frac{1}{\sqrt{2}} & 0 & 0 & 0 & 0 \\
0 & 0 & \frac{1}{\sqrt{2}} & -\frac{1}{\sqrt{2}} & 0 & 0 & 0 & 0 \\
\hline
0 & 0 & 0 & 0 & \frac{1}{\sqrt{2}} & \frac{1}{\sqrt{2}} & 0 & 0 \\
0 & 0 & 0 & 0 & \frac{1}{\sqrt{2}} & -\frac{1}{\sqrt{2}} & 0 & 0 \\
\hline
0 & 0 & 0 & 0 & 0 & 0 & \frac{1}{\sqrt{2}} & \frac{1}{\sqrt{2}} \\
0 & 0 & 0 & 0 & 0 & 0 & \frac{1}{\sqrt{2}} & -\frac{1}{\sqrt{2}} \\
\end{array} \right) \quad . 
\end{eqnarray}

With the matrix $\tilde{M}[\hat{H}]$ given as above, the full transformation of the simplex vector is given by Eq.~(14), i.e., $T[\hat{H}](\vec{s}) =  \frac{1}{8} \left( \tilde{I} - \tilde{M}[\hat{H}] \right) \cdot \vec{u} + \tilde{M}[\hat{H}] \cdot \vec{s}$. As we discussed in detail in our recent paper \cite{deniz_rudhy}, the above analysis can be extended to find the transformation matrices for Rabi rotations as well as the single-qubit phase gate. For completeness, we present these matrices in Appendix A. Because any arbitrary evolution of the single-qubit wavefunction can be achieved using a combination of Rabi rotation gate and phase-gates, such evolution can be fully tracked using corresponding transformations of the corresponding vector $\vec{s}$ in the probability space (see Fig. 8 in Appendix A for more clarity). 

\subsection{Measurements on the single-qubit system}

As we mentioned above, the above map of the single-qubit wavefunction stores both the real and imaginary parts of the complex coefficients in an eight-dimensional vector of probabilities $\vec{s}$. Not surprisingly, there is also a one-to-one correspondence between the measurement outcomes. That is, by measuring the probabilities of the mapped system (i.e., the components of the vector $\vec{s}$), we can calculate the probabilistic outcomes of the measurement of the quantum system.

With the complex coefficients given in the qubit wavefunction, $c_0$ and $c_1$, when a quantum measurement is performed, the corresponding probabilities of finding the system in state $\ket{0}$ and $\ket{1}$ are $\vert c_0 \vert^2$ and $\vert c_1 \vert^2$, respectively. These quantities, in turn, can be expressed as the probability components of the simplex vector: 
\begin{eqnarray}
\vert c_0 \vert^2 = x_0^2 + y_0^2= (1-8s_1)^2 + (1-8s_5)^2 \quad , \nonumber \\
\vert c_1 \vert^2 = x_1^2 + y_1^2= (1-8s_2)^2 + (1-8s_6)^2 \quad .
\end{eqnarray}

More generally, we can also establish a one-to-one correspondence between measurements of an observable in the quantum system and corresponding measurements in the probability space. For a given observable $\hat{A}$ and a quantum state $\ket{\psi}$, the average measured value of $\hat{A}$ when the quantum system is in state $\ket{\psi}$ is $\langle \hat{A}\rangle_{\ket{\psi}}= \bra{\psi}\hat{A}\ket{\psi}$. To establish a corresponding quantity in the probability space, we first map the operator $\hat{A}$ in an identical way to how we mapped the evolution operator, $\hat{U}$, above:
\begin{equation}
	\tilde{M}[\hat{A}]=\left(\begin{array}{c| c| c| c}
		\Re(\hat{A})&O&O&\Im(\hat{A})\\
        \hline
		O&\Re(\hat{A})&\Im(\hat{A})&O\\
        \hline
		\Im(\hat{A})&O&\Re(\hat{A})&O\\
        \hline
		O&\Im(\hat{A})&O&\Re(\hat{A})
	\end{array}\right)=\tilde{I}_{4\times4}\otimes\Re(\hat{A}) + \tilde{\Lambda}\otimes\Im(\hat{A}) \quad ,
\end{equation}
It can then be shown that, the same average measurement value for the quantum observable can be obtained by the following expression in terms of the mapped observable, $\tilde{M}[\hat{A}]$, and mapped quantum state, $\varphi\ket{\psi}=\vec{s} = (\vec{u}+\vec{p})/8$. 
\begin{equation}
   \bra{\psi}\hat{A}\ket{\psi} =  (\vec{p}^{\mathsf{T}}\cdot\tilde{M}[\hat{A}]\cdot\vec{p})/2 \quad . 
\end{equation}
We prove this correspondence in Appendix B. We have found that there is another informative way to evaluate the average measured value of the quantum observable $\hat{A}$ in the probability space. We can envision to evolve the mapped simplex vector, $\vec{s}$, with the transformation matrix of $\tilde{M}[\hat{A}]$ of above, which we refer to as $T[\hat{A}](\vec{s})$.  We can then estimate how much this transformed vector has deviated from the initial simplex vector, which we denote with $\langle T[\hat{A}]\rangle_{\vec{s}}$, by taking the dot product of the transformed vector with the initial one:
\begin{equation}
    \langle T[\hat{A}]\rangle_{\vec{s}}\equiv \vec{s}^{\mathsf{T}}\cdot T[\hat{A}](\vec{s})
\end{equation}
We can then relate this quantity to the quantum measurement outcome, $\langle \hat{A}\rangle_{\ket{\psi}}$, by noting that:
\begin{equation}
    \vec{s}^{\mathsf{T}}\cdot T[\hat{A}](\vec{s})=\frac{1}{8}(\vec{s}^{\mathsf{T}}\cdot\vec{u}+\vec{s}^{\mathsf{T}}\cdot\tilde{M}[\hat{A}]\cdot\vec{p})=\frac{1}{8}[1+\frac{1}{8}(\vec{u}^{\mathsf{T}}+\vec{p}^{\mathsf{T}})\cdot\tilde{M}[\hat{A}]\cdot\vec{p}]=\frac{1}{8}[1+\frac{1}{8}(\vec{p}^{\mathsf{T}}\cdot\tilde{M}[\hat{A}]\cdot\vec{p})]
\end{equation}
which then gives:
\begin{equation}
    \langle T[\hat{A}]\rangle_{\varphi\ket{\psi}}=\frac{1}{8}(1+\frac{1}{4}\langle\hat{A}\rangle_{\ket{\psi}}) \quad . 
\end{equation}
This equation forms a direct correspondence between quantum measurements and measurements performed in the probability space, for any quantum observable $\hat{A}$.


\section{IV. Extension to multiple qubits}
\subsection{Mapping the wavefunction of two-qubits}
When we have more than one qubit, the Hilbert space is given by the tensor product of the Hilbert space of the individual qubits; i.e., for two qubits, the state will be of the form:
\begin{eqnarray}
|\psi \rangle = |\psi_1 \rangle \otimes |\psi_2 \rangle \quad .  
\end{eqnarray}
For probabilistic spaces, we combine multiple vectors in an identical way. This has been discussed and rigorously proven in Ref.~\cite{barrett1}; it is also implicit in the discussion of the mathematical structure of the probability theory by de Finetti \cite{definetti}. However, this feature of combining probability spaces is not widely known. It is usually assumed that the tensor product is a feature that is special and specific to quantum mechanics. 

To map two qubits, we first envision mapping each qubit to a simplex vector, exactly as defined above with the single qubit map: $\varphi\ket{\psi_1} = \vec{s}_1$ and $\varphi\ket{\psi_2} = \vec{s}_2$. The combined vector in the simplex will be given by
\begin{eqnarray}
\vec{s}'_{12} = \vec{s}_1 \otimes \vec{s}_2 \quad . 
\end{eqnarray}
In this combined vector, we again have something quite physical in mind. Instead of one, we now have two ``dice", each with eight probabilistic outcomes. The 64 entries of the vector $\vec{s}'_{12}$ stores the joint probability distribution of two ``dice" experiment outcomes. 

We note that this form of the combined vector is not of the form of the single-qubit simplex vector as described above by Eq.~(3). Specifically, looking at this combined vector more closely:
\begin{eqnarray}
    \vec{s}'_{12} & = & \vec{s}_1 \otimes \vec{s}_2 \quad \nonumber \\ & = & \frac{1}{8^2} (\vec{u} + \vec{p}_1) \otimes (\vec{u} + \vec{p}_2) \quad \nonumber \\
    & =  & \frac{1}{8^2}(\vec{u}\otimes\vec{u}+\vec{u}\otimes\vec{p}_2+\vec{p}_1\otimes\vec{u}+\vec{p}_1\otimes\vec{p}_2)\quad .
\end{eqnarray}
There are two cross-terms on the right hand side of Eq.~(36), $\vec{u}\otimes\vec{p}_2+\vec{p}_1\otimes\vec{u}$, which prevents the combined vector $\vec{s}'_{12}$,  taking the form of the simplex vector as defined by Eq.~(3). As will be clear below, to be able to extend unitary operations for the two-qubits to transformations in the overall simplex vector (i.e.,  to extend transformation matrices $\tilde{M}$ to the combined simplex vector), it is imperative that we retain the form in Eq.~(3) for the combined vector. We require a joint probability distribution, which we define $\vec{s}_{12}$, containing information of the two simplex vectors, of the form, 
\begin{equation}
    \vec{s}_{12}=\frac{1}{8^2}(\vec{u}\otimes\vec{u}+\vec{p}_1\otimes\vec{p}_2) \quad .
\end{equation}
This joint distribution, $\vec{s}_{12}$ is of the form Eq.~(3), and does not have the cross terms. As we will discuss below in detail, for this joint distribution, $\vec{s}_{12}$, quantum unitary operations can now be formulated by taking tensor product of operations on $\vec{p}_1$ and $\vec{p}_2$ with some final offset (i.e., affine operations), analogous to the procedure for the transformations mimicking single-qubit gates that we discussed above. 
\begin{figure}[h]
    \centering
    \includegraphics[width=0.4\textwidth]{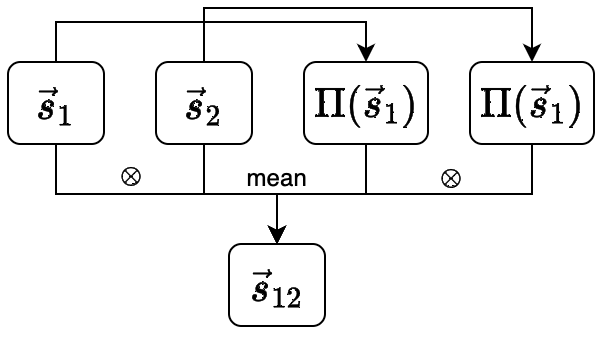}
    \caption{Simplified schematic for the affine transformation that generates the $\vec{s}_{12}$ state with the required form, starting with the initial states $\vec{s}_1$ and $\vec{s}_2$. The $\otimes$ symbol represents tensor product of the corresponding two states and ``mean" here implies that we add the resultant tensored vectors and divide by $2$ as in Eq.~(41) (i.e., taking a statistical average).}
\end{figure}

The procedure for obtaining the combined vector $\vec{s}_{12}$ of the desired form is as follows. By applying appropriate affine transformations to the vectors, $\vec{s}_1$ and $\vec{s}_2$, we can also generate a joint distribution that has the opposite signs of the above-mentioned cross terms. We then take a statistical mixture of $\vec{s}'_{12}$, with its copy that has the cross terms with the opposite signs. More formally, we use the following bi-affine transformation, $\tau$: 
\begin{equation}
    \vec{s}_{12}= \tau(\vec{s}_1, \vec{s}_2) \equiv \frac{1}{2}\left[\vec{s}_1\otimes\vec{s}_2+\Pi(\vec{s}_1)\otimes\Pi(\vec{s}_2)\right]
\end{equation}
Here, the transformations of the single vectors $\Pi(\vec{s}_1)$ and $\Pi(\vec{s}_2)$ are the following: 
\begin{eqnarray}
    \Pi(\vec{s}_1) & = & \frac{1}{8}(\vec{u}+ \tilde{\Pi} \cdot \vec{p}_1)= \frac{1}{8}(\vec{u}-\vec{p}_1)  \quad , \nonumber \\ 
    \Pi(\vec{s}_2) & = & \frac{1}{8}(\vec{u}+ \tilde{\Pi}  \cdot \vec{p}_2)= \frac{1}{8}(\vec{u}-\vec{p}_2) \quad .
\end{eqnarray}
In above, the matrix $\tilde{\Pi}$ is a projection matrix which shuffles the entries of the $\vec{p}$ vector to map $\vec{p} \rightarrow -\vec{p}$. We note that the transformations $\Pi(\vec{s})$ are affine since for any two $\lambda$ and $\lambda'$ such that $\lambda + \lambda' =1 $, and any two simplex vectors $\vec{s}$ and $\vec{s}'$, we have $\Pi(\lambda \vec{s} + \lambda' \vec{s}') = \lambda \Pi (\vec{s}) + \lambda' \Pi( \vec{s}')$. Because the transformation $\Pi(\vec{s})$ is affine, the transformation of the combined simplex vector, $\tau(\vec{s}_1, \vec{s}_2)$ is also affine in each of its entries. That is, for any two constants $\lambda$ and $\lambda'$ such that $\lambda + \lambda' =1 $ and simplex vectors, $\vec{s}_1$,  $\vec{s}_1'$, and $\vec{s}_2$, $\vec{s}_2'$, we have:
\begin{eqnarray}
\tau(\lambda \vec{s}_1 + \lambda' \vec{s}_1', \vec{s}_2 ) = \lambda \tau(\vec{s}_1, \vec{s}_2) + \lambda' \tau(\vec{s}_1', \vec{s}_2) \quad , \nonumber \\
\tau(\vec{s}_1, \lambda \vec{s}_2 + \lambda' \vec{s}_2') = \lambda \tau(\vec{s}_1, \vec{s}_2) + \lambda' \tau(\vec{s}_1, \vec{s}_2') \quad .
\end{eqnarray}
The bi-affine transformation of Eq.~(38) will be the starting point for mapping and manipulation of multiple qubits and will be used throughout the manuscript continually. Because of this, we define a new operation which we call $\sotimes$, which essentially refers to mapping of multiple qubits followed by this unique transformation to eliminate the cross terms:
\begin{equation}
    \varphi_2(\ket{\psi_1}\otimes\ket{\psi_2}) \equiv \vec{s}_1\sotimes\vec{s}_2 \equiv \tau(\vec{s}_1, \vec{s}_2) = \frac{1}{2}\left[\vec{s}_1\otimes\vec{s}_2+\Pi(\vec{s}_1)\otimes\Pi(\vec{s}_2)\right] \quad . 
\end{equation}
Below in Appendix C, we will take a closer look at this bivalent operation; we will define it more rigorously and show that it is  closed and satisfies all the properties of the ordinary tensor operation.

Below, in all the quantum algorithms that we discuss, an initial system of unentangled qubits (i.e., with the $n$-qubit wavefunction in a product state), will initially mapped to a tensor product of simplex vectors, in exactly the same manner as we described above. We will then apply a transformation similar to Eq.~(41), to transform the overall simplex vector in a form similar to $\vec{s}_{12}$ of Eq.~(38) (i.e., in a form which is a constant added to the tensor product of individual $\vec{p}$ vectors). We will then show how a sequence of single-qubit and two-qubit gates that evolve the wavefunction can be mimicked in the probability space, with the overall simplex vector smoothly following the wavefunction. 

We note that the map that is shown in Eq.~(41) can also be extended to initial non-separable two-qubit states. A general two-qubit state wavefunction can be written as a linear combination of separable tensor product states: $\ket{\psi} =\sum_{j,k}\ket{\psi_j}\otimes\ket{\psi_k} $. With each state $\ket{\psi_j}$ producing coefficients in the simplex vector $\vec{p}_j$, and since the map $\varphi_2$ is linear for the $\vec{p}$ vectors we can define,
\begin{equation}
    \varphi_2(\sum_{j,k}\ket{\psi_j}\otimes\ket{\psi_k})\equiv\frac{1}{8^2}(\vec{u}^{\otimes_2}+\sum_{j,k}\vec{p}_j\otimes\vec{p}_k) \quad . 
\end{equation}

One interesting property of the map of Eq.~(3) is that, the absolute phase of the quantum wavefunction matters. That is, the quantum states $\ket{\psi}$ and $\exp{i \phi} \ket{\psi}$ are mapped to different vectors $\vec{s}$ in the probability space. A consequence of this is that, when multiple qubits are mapped, the absolute phases of each qubit wavefunction cannot be trivially combined, and the ordering of these phases become important. Such ordering of the phases will be important in the Quantum Fourier Transform discussion of below and we will also  discuss it more thoroughly in Appendix D.

\subsection{Implementing two-qubit gates}
In this section, we discuss the implementation of operations on the two qubits using appropriate transformations of the mapped and transformed simplex vector, $\vec{s}_{12}$. There are two types of operations that we will consider: (1) separable operations where two independent single-qubit gates are applied to each qubit, and (2) non-separable operations such as the entangling two-qubit controlled-not (CNOT) gate. 

 We start with separable operations on the two qubits of the form $\hat{U}=\hat{U}_1\otimes \hat{U}_2$, where the first qubit and second qubit each evolve under operators $\hat{U}_1$ and $\hat{U}_2$, respectively. In simplex space, we define the tensor product of operations on each of the individual simplex vectors: $\tilde{M}_2[\hat{U}]\equiv\tilde{M} [\hat{U}_1]\otimes\tilde{M}[\hat{U}_2]$. More explicitly, $\tilde{M}_2[\hat{U}]$ is $64 \times 64 $ matrix, which is a tensor product of two $8 \times 8 $ single simplex vector transformation matrices as we discussed above:  
\begin{equation}
    \tilde{M}_2[\hat{U}_1\otimes \hat{U}_2]\doteq\left(\begin{array}{c| c| c| c}
		\Re(\hat{U}_1)&O&O&\Im(\hat{U}_1)\\
        \hline
		O&\Re(\hat{U}_1)&\Im(\hat{U}_1)&O\\
        \hline
		\Im(\hat{U}_1)&O&\Re(\hat{U}_1)&O\\
        \hline
		O&\Im(\hat{U}_1)&O&\Re(\hat{U}_1)
	\end{array}\right)\otimes\left(\begin{array}{c| c| c| c}
		\Re(\hat{U}_2)&O&O&\Im(\hat{U}_2)\\
        \hline
		O&\Re(\hat{U}_2)&\Im(\hat{U}_2)&O\\
        \hline
		\Im(\hat{U}_2)&O&\Re(\hat{U}_2)&O\\
        \hline
		O&\Im(\hat{U}_2)&O&\Re(\hat{U}_2)
	\end{array}\right) \quad . 
\end{equation}
We apply the transformation matrix of above, to the combined simplex vector, $\vec{s}_{12}$ exactly in the same manner as the single simplex vector transformations:
\begin{equation}
    T_2[\hat{U}](\vec{s}_{12})\equiv\frac{1}{8^2}(\tilde{I}_{64\times64}-\tilde{M}_2[\hat{U}])\cdot\vec{u}\otimes\vec{u}+\tilde{M}_2[\hat{U}]\cdot\vec{s}_{12}\quad .
\end{equation}
Because the form of the combined simplex vector, $\vec{s}_{12}$,  is identical to the single simplex vector of Eq.~(3), the above transformation of $\vec{s}_{12}$ results in appropriate linear transformations of the corresponding $\vec{p}$ vectors. More explicitly, we have
\begin{align}
    T_2[\hat{U}](\vec{s}_{12})&=\frac{1}{8^2}(\tilde{I}_{64\times64}-\tilde{M}[\hat{U}])\cdot\vec{u}\otimes\vec{u}+\frac{1}{8^2}\tilde{M}[\hat{U}]\cdot(\vec{u}\otimes\vec{u}+\vec{p}_1\otimes\vec{p}_2)\\
    &=\frac{1}{8^2}(\vec{u}\otimes\vec{u}+\tilde{M}[\hat{U}_1]\cdot\vec{p}_1\otimes\tilde{M}[\hat{U}_2]\cdot\vec{p}_2)
\end{align}
We note that, when $\tilde{M}_2[\hat{U}]\equiv\tilde{M} [\hat{U}_1]\otimes\tilde{M}[\hat{U}_2]$ of Eq.~(43) is explicitly evaluated, the final matrix would contain terms with all four product combinations of the real and imaginary parts of the individual evolution operators, i.e., $\Re(\hat{U}_1) \otimes \Re(\hat{U}_2)$, $\Re(\hat{U}_1) \otimes \Im(\hat{U}_2)$, $\Im(\hat{U}_1) \otimes \Re(\hat{U}_2)$, and $\Im(\hat{U}_1) \otimes \Im(\hat{U}_2)$. As a result, we cannot express $ \tilde{M} [\hat{U}_1]\otimes\tilde{M}[\hat{U}_2]$ just in terms of $\Re(\hat{U}_1\otimes \hat{U}_2)$ and $\Im(\hat{U}_1\otimes \hat{U}_2) $. Hence we have the following important inequality:
\begin{equation}
    \tilde{M}[\hat{U}_1\otimes \hat{U}_2]\neq \tilde{I}_{4\times4}\otimes \Re(\hat{U}_1\otimes \hat{U}_2)+\Lambda\otimes \Im(\hat{U}_1\otimes \hat{U}_2) \quad . 
\end{equation}
We, therefore, conclude that the definition of $\tilde{M}$ (single qubit operator map) is different from $\tilde{M}_2$ (two qubit operator map). As we will discuss below, for multiple qubit separable operations $\hat{U}=\bigotimes_{k=1}^{n}\hat{U}_k$ the map is analogously defined as $\tilde{M}_n[\hat{U}]\equiv\bigotimes_{k=1}^{n}\tilde{M}[\hat{U}_k]$.

We now move to non-separable, entangling operations, such as the CNOT gate. A general two qubit controlled unitary operation $\hat{C}_{U}$ can be decomposed as:
\begin{equation}
    \hat{C}_{\hat{U}}=\hat{P}_0\otimes \hat{I}_{2\times2}+\hat{P}_1\otimes \hat{U},\quad P_{0}=\op{0},\quad P_{1}=\op{1}
\end{equation}
Here, the first qubit is the control qubit, and the second qubit is the target qubit, respectively. The above unitary operation $\hat{C}_{\hat{U}}$ applies identity operator to the target qubit, when the control qubit is in state $\ket{0}$. When the control qubit is in state $\ket{1}$, the target qubit evolves under unitary operator $\hat{U}$. Based on the above unitary controlled quantum operation,  we define its simplex analog $\tilde{M}_2[\hat{C}_{\hat{U}}]$ as,
\begin{equation}
    \tilde{M}_2[\hat{C}_{\hat{U}}]\equiv\tilde{P}_{0}\otimes \tilde{I}_{8\times8}+\tilde{P}_1\otimes\tilde{M}[\hat{U}]
\end{equation}
where the two ``projection" matrices, $\tilde{P}_0$ and $\tilde{P}_1$, acting on the first simplex vector are 
\begin{eqnarray}
\tilde{P}_0= \left( \begin{array}{cc|cc|cc|cc}
1 & 0 & 0 & 0 &  0 & 0 & 0 & 0 \\
0 & 0 & 0 & 0 &  0 & 0 & 0 & 0 \\
\hline
0 & 0 & 1 & 0 &  0 & 0 & 0 & 0 \\
0 & 0 & 0 & 0 &  0 & 0 & 0 & 0 \\
\hline
0 & 0 & 0 & 0 &  1 & 0 & 0 & 0 \\
0 & 0 & 0 & 0 &  0 & 0 & 0 & 0 \\
\hline
0 & 0 & 0 & 0 &  0 & 0 & 1 & 0 \\
0 & 0 & 0 & 0 &  0 & 0 & 0 & 0 \\
\end{array} \right) \quad 
\tilde{P}_1= \left( \begin{array}{cc|cc|cc|cc}
0 & 0 & 0 & 0 &  0 & 0 & 0 & 0 \\
0 & 1 & 0 & 0 &  0 & 0 & 0 & 0 \\
\hline
0 & 0 & 0 & 0 &  0 & 0 & 0 & 0 \\
0 & 0 & 0 & 1 &  0 & 0 & 0 & 0 \\
\hline
0 & 0 & 0 & 0 &  0 & 0 & 0 & 0 \\
0 & 0 & 0 & 0 &  0 & 1 & 0 & 0 \\
\hline
0 & 0 & 0 & 0 &  0 & 0 & 0 & 0 \\
0 & 0 & 0 & 0 &  0 & 0 & 0 & 1 \\
\end{array} \right) \quad . 
\end{eqnarray}
With the transformation matrix, $\tilde{M}_2[\hat{C}_{\hat{U}}]$ described by Eq.~(49), we transform the combined simplex vector $\vec{s}_{12}$ exactly in the manner described by Eq.~(33): $ \vec{s}_{12} \longrightarrow \frac{1}{8^2}(\tilde{I}_{64\times64}-\tilde{M}_2[\hat{C}_{\hat{U}}])\cdot\vec{u}\otimes\vec{u}+\tilde{M}_2[\hat{C}_{\hat{U}}]\cdot\vec{s}_{12}$. We note that this transformation is a valid transformation of the simplex vector since 
(1) the operation $\hat{C}_{\hat{U}}$ is unitary as a whole implying that $T_2[\hat{C}_{\hat{U}}]$ is affine and conserves probability under transformations, (2) we can explicitly prove that $\tilde{M}_2[\hat{C}_{\hat{U}}]\cdot\vec{P}$ from some valid $\vec{P}$ always has the form such that it is orthogonal to $\vec{u}\otimes\vec{u}$ as was suggested earlier in Eq.~(2). Let $\vec{P}=\sum_{jk}\vec{p}_{j}\otimes\vec{p}_k$, then,
\begin{align*}
    \nonumber\vec{u}^{\mathsf{T}}\otimes\vec{u}^{\mathsf{T}}\cdot(\tilde{M}[\hat{C}_{\hat{U}}]\cdot\vec{P})&=\sum_{j,k}(\vec{u}^{\mathsf{T}}\otimes\vec{u}^{\mathsf{T}}\cdot(\tilde{P}_{0}\otimes \tilde{I}_{8\times8})\cdot(\vec{p}_j\otimes\vec{p}_k)+\vec{u}^{\mathsf{T}}\otimes\vec{u}^{\mathsf{T}}\cdot(\tilde{P}_1\otimes\tilde{M}(\hat{U}))\cdot(\vec{p}_j\otimes\vec{p}_k))\\
    &=\sum_{j,k}\left((\vec{u}^{\mathsf{T}}\cdot\tilde{P}_0\cdot\vec{p}_j)\cancelto{0}{(\vec{u}^{\mathsf{T}}\cdot\vec{p}_k)}+(\vec{u}^{\mathsf{T}}\cdot\tilde{P}_1\cdot\vec{p}_j)\cancelto{0}{(\vec{u}^{\mathsf{T}}\cdot\tilde{M}[\hat{U}]\cdot\vec{p}_k)}\quad\right)=0
\end{align*}
\begin{figure}
    \centering
    \includegraphics[width=0.5\textwidth]{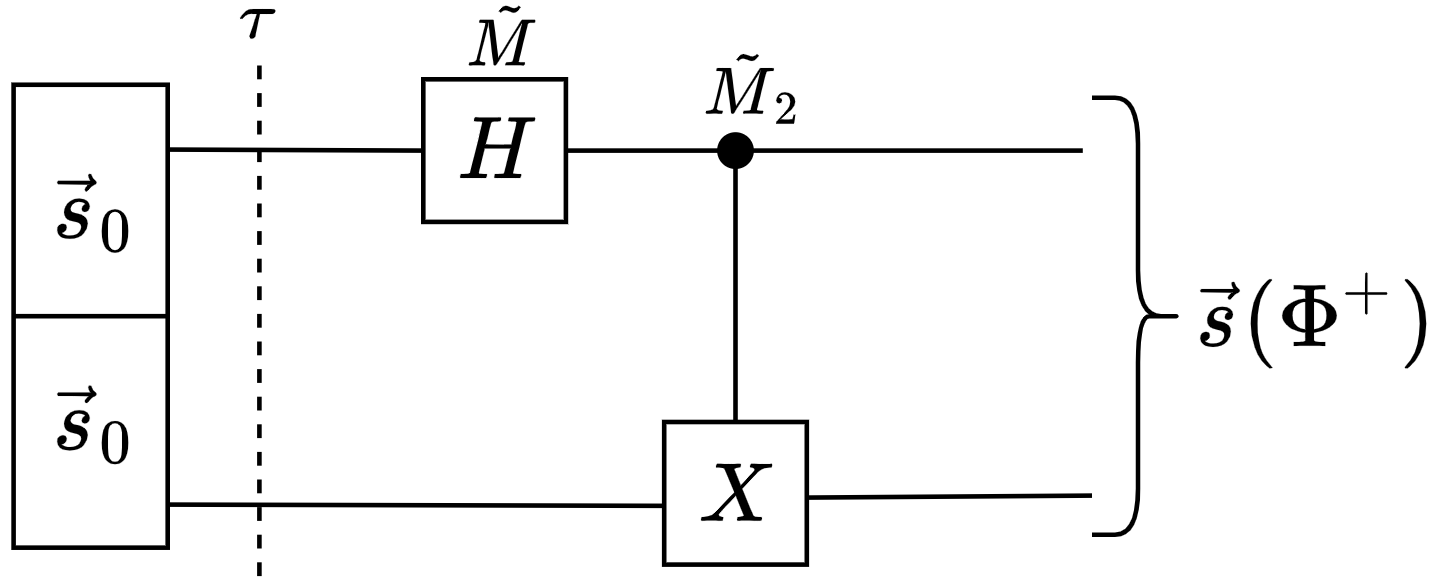}
    \caption{The circuit for creating a Bell state in the probability space. The dashed line marked as $\tau$ indicates the application of the bi-affine map $\tau$ (see Eq.~(38)) to the initial state, i.e. the state after the dashed line in this case will be: $\tau(\vec{s}_0,\vec{s}_0)=\vec{s}_{00}$ as defined in Eq.~(51).}
\end{figure}
In order to make these ideas more concrete,  we next consider the creation of the simplex version $\vec{s}(\Phi^+)$ of the two-qubit entangled Bell state $\ket{\Phi^+}$. The simplex circuit that accomplishes this is shown schematically in Fig.~3. Here, we start with the $\vec{s}_0\otimes\vec{s}_0$ state (exactly mimicking a two-qubit quantum system starting in the $\ket{00}$ state) and use the above introduced map $\tau$ to produce the $\vec{s}_{00}$ state,
\begin{equation}
    \vec{s}_{00}\equiv\vec{s}_0\sotimes\vec{s}_0= \tau(\vec{s}_0, \vec{s}_0) = \frac{1}{8^2}(\vec{u}\otimes\vec{u}+\vec{p}_0\otimes\vec{p}_0) \quad .
\end{equation}
We then apply the analog of the Hadamard gate to the first simplex vector transforming that vector into:
\begin{align}
    \vec{s}'=T_2[\hat{H}\otimes \hat{I}](\vec{s}_{00})&=\frac{1}{8^2}(\vec{u}\otimes\vec{u}+\tilde{M}[\hat{H}]\cdot\vec{p}_0\otimes \tilde{I}_{8\times8}\cdot\vec{p}_0)\\
    &=\frac{1}{8^2}[\vec{u}\otimes\vec{u}+(\vec{p}_0(1/\sqrt{2})+\vec{p}_1(1/\sqrt{2}))\otimes\vec{p}_0]\\
    &=\frac{1}{8^2}[\vec{u}\otimes\vec{u}+\frac{1}{\sqrt{2}}(\vec{p}_0+\vec{p}_1)\otimes\vec{p}_0]=\frac{1}{8^2}[\vec{u}\otimes\vec{u}+\frac{1}{\sqrt{2}}(\vec{p}_{00}+\vec{p}_{10})] \quad . 
\end{align}
Here, we have introduced the following useful notation, which we will use extensively throughout the manuscript
\begin{eqnarray}
\vec{p}_{00} \equiv \vec{p}_0 \otimes \vec{p}_0 \quad , \quad  \vec{p}_{01} \equiv \vec{p}_0 \otimes \vec{p}_1 \quad , \nonumber \\
\vec{p}_{10} \equiv \vec{p}_1 \otimes \vec{p}_0 \quad , \quad  \vec{p}_{11} \equiv \vec{p}_1 \otimes \vec{p}_1 \quad .
\end{eqnarray}
More generally, for $n$ qubits, we will use an extended version of this notation:
\begin{equation}
    \bigotimes_{i=1}^{n}\vec{p}_{q_i}\equiv \vec{p}_{\bf{q}},\quad {\bf{q}} = (q_1,\dots,q_n) \in \mathbb{B}^n\quad .
\end{equation}
After the equivalent of the Hadamard gate on the first vector, we next apply the analog of the CNOT gate to obtain the Bell-state in the probability space:
\begin{align}
    \vec{s}(\Phi^+)&=T[C_{\hat{X}}](\vec{s}')=\frac{1}{8^2}[\vec{u}\otimes\vec{u}+\tilde{M}_2[\hat{C}_{\hat{X}}]\cdot\frac{1}{\sqrt{2}}(\vec{p}_{00}+\vec{p}_{10})]\\
    &=\frac{1}{8^2}[\vec{u}\otimes\vec{u}+\frac{1}{\sqrt{2}}(\tilde{P}_{0}\otimes \tilde{I}_{8\times8}+\tilde{P}_1\otimes\tilde{M}_2[\hat{X}])\cdot(\vec{p}_{0}\otimes\vec{p}_0+\vec{p}_{1}\otimes\vec{p}_0)]\\
    &=\frac{1}{8^2}[\vec{u}\otimes\vec{u}+\frac{1}{\sqrt{2}}(\vec{p}_{0}\otimes\vec{p}_0+\vec{p}_{1}\otimes\vec{p}_1)]\\
    &=\frac{1}{8^2}(\vec{u}\otimes\vec{u}+\frac{1}{\sqrt{2}}(\vec{p}_{00}+\vec{p}_{11}))
\end{align}
hence we see that the state $\vec{s}(\Phi^+)$ has the form that exactly mimicks the quantum system being in a two-qubit entangled Bell state of the form, $\ket{\psi} = \frac{1}{\sqrt{2}} ( \ket{00} + \ket{11} )$.

\subsection{Mapping states and operators for more than two qubits}
Extending the map for more than two qubits is straightforward. We first view each qubit to be mapped to a corresponding eight dimensional $\vec{s}$ vector, exactly in the manner described by Eq.~(3). The tensor product of the qubit wavefunctions would then map to the tensor product of probability vectors, $\vec{s}_1 \otimes \vec{s}_2 \cdots \otimes \vec{s}_n$. However, identical to the two-qubit case, this would produce cross terms in the final resultant vector. We then use the procedure outlined above and use simplex tensor operation $\sotimes$ to obtain
\begin{equation}
\varphi_{n}\left(\bigotimes_{k=1}^{n}\ket{\psi_k}\right)\equiv\vec{s}_{1}\sotimes\vec{s}_{2}\,\cdots\,\sotimes\vec{s}_{n}=\frac{1}{8^n}(\vec{u}^{\otimes^{n}}+\bigotimes_{k=1}^{n}\vec{p}_{k})=\vec{s}_{1,n} \quad . 
\end{equation}
\begin{figure}
    \centering
    \includegraphics[width=0.8\textwidth]{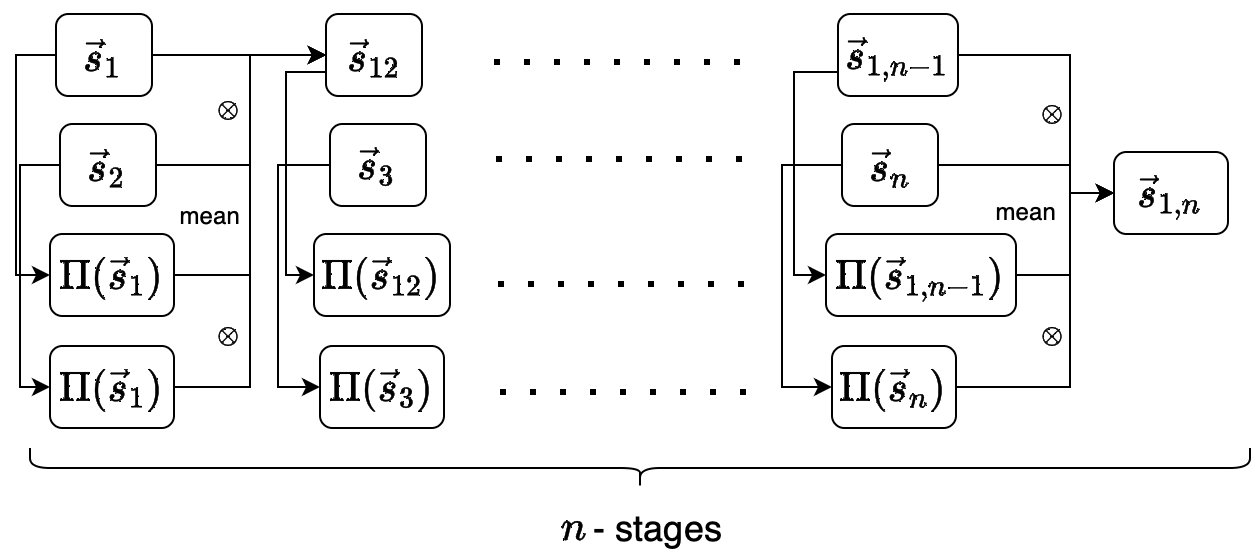}
    \caption{Simplified schematic for the operations involved in the recursive application of the bi-affine map $\tau$ as defined in Eq.~(62) (cf. Fig. 2 and refer to Appendix C for details). At each stage we generate a new distribution using the bi-affine map $\tau$ and propagate that distribution to the next stage. After $n$ stages, the circuit produces a combined simplex state $\vec{s}_{1,n}$ corresponding to the $n$ qubit state. From this diagram it is clear that the runtime/time complexity of $\varphi_n$ grows linearly with $n$ because there are only $n$-stages required with each costing a constant overhead.}
\end{figure}
This can be achieved by recursive application of the affine map $\tau$ introduced earlier, in the following manner: 
\begin{equation}
\varphi_{n}\left(\bigotimes_{k=1}^{n}\ket{\psi_k}\right)\equiv\tau(\vec{s}_1,\tau(\vec{s}_{2},\dots\tau(\vec{s}_{n-1},\vec{s}_n)\dots)) \quad . 
\end{equation}
We provide a schematic for the circuit that achieves the recursive application of this affine map in Fig.~4. The overall map $\varphi_n$ is as before also affine for each of the simplex state vectors. We note that since the action of each bi-affine transformation $\tau$ can be applied in constant time and memory, we can implement the map $\varphi_n$ (i.e, the circuit that is shown in Fig.~4) for any large $n$ in $O(n)$ time and memory. For non-separable states the map is defined based on Eq.~(42) as, 
\begin{equation}
    \varphi_{n}\left(\sum_{j}\bigotimes_{i=1}^{n}\ket{\psi_{j_i}}\right)=\frac{1}{8^n}(\vec{u}^{\otimes^n}+\sum_{j}\bigotimes_{i=1}^{n}\vec{p}_{j_i})
\end{equation}
where, each $\vec{p}_{j_i}$ correspond to the $p$-state of each $\varphi\ket{\psi_{j_i}}$.
For mapping unitary evolution, first we consider separable (non-entangling) operators of the form $\bigotimes_{i=1}^{n}\hat{U}_{i}$, which is a tensor product of evolution of each qubit by operator $\hat{U}_i$, respectively. Following exactly the same strategy as we discussed above for two-qubits, we define the map of this tensor product as a tensor product of individual mapped operators:
\begin{equation}    \tilde{M}_n[\bigotimes_{i=1}^{n}\hat{U}_{i}]\equiv \bigotimes_{i=1}^{n}\tilde{M}[\hat{U}_{i}] \quad .
\end{equation}
Here, the map for each unitary operator acting on qubit $i$, $\hat{U}_i$, is exactly as defined above:
\begin{equation}
	\tilde{M}[\hat{U}_i]=\left(\begin{array}{c| c| c| c}
		\Re(\hat{U}_i)&O&O&\Im(\hat{U}_i)\\
        \hline
		O&\Re(\hat{U}_i)&\Im(\hat{U}_i)&O\\
        \hline
		\Im(\hat{U}_i)&O&\Re(\hat{U}_i)&O\\
        \hline
		O&\Im(\hat{U}_i)&O&\Re(\hat{U}_i)
	\end{array}\right)=\tilde{I}_{4\times4}\otimes\Re(\hat{U}_i) + \tilde{\Lambda}\otimes\Im(\hat{U}_i) \quad .
\end{equation}
More generally, entangling (non-separable) operations can be expressed as a sum of separable operations of the form $\sum_{j}\bigotimes_{i=1}^{n}\hat{U}_{j_i}$, and their map to the probability space is:
\begin{equation}
\tilde{M}_n[\sum_{j}\bigotimes_{i=1}^{n}\hat{U}_{j_i}]\equiv\sum_{j}\bigotimes_{i=1}^{n}\tilde{M}[\hat{U}_{j_i}] \quad . 
\end{equation}
The transformation of the overall simplex vector with this mapped evolution is:
\begin{align}
    T_{n}[\sum_{j}\bigotimes_{i=1}^{n}\hat{U}_{j_i}](\vec{S})&=\frac{1}{8^n}(\tilde{I}_{8^n\times8^n}-\tilde{M}_n[\sum_{j}\bigotimes_{i=1}^{n}\hat{U}_{j_i}])\cdot\vec{u}^{\otimes^n}+\tilde{M}_n[\sum_{j}\bigotimes_{i=1}^{n}\hat{U}_{j_i}]\cdot\vec{S}\\
    &=\frac{1}{8^n}(\tilde{I}_{8^n\times8^n}-\sum_{j}\bigotimes_{i=1}^{n}\tilde{M}[\hat{U}_{j_i}])\cdot\vec{u}^{\otimes^n}+\sum_{j}\bigotimes_{i=1}^{n}\tilde{M}[\hat{U}_{j_i}]\cdot\vec{S} \quad , 
\end{align}
for some given mapped $n$-qubit simplex state vector $\vec{S}$.

\subsection{Measurements on the $n$-qubit system}
The map defined above for $n$ qubits stores the real and imaginary parts of the complex coefficients in the corresponding probabilities of the $\vec{s}$ vector. Identical to the single qubit maps that we discussed above, because the real and imaginary parts are stored in the mapped simplex vector, there is one-to-one correspondence between the measurement probabilities of the quantum wavefunction and the measured individual entries (i.e., probabilities) of the simplex vector. Following the strategy introduced for single qubit measurements, for a given $n$-qubit total wavefunction $\ket{\psi_{tot}}$, the corresponding mapped simplex vector $\vec{s}_{tot}=\varphi_n\ket{\psi_{tot}}$ and the qubit projection operator $\hat{M}_{\bf{q}}=\op{\bf{q}}$, the following connection holds (cf. Eq~(33)):
\begin{equation}
    \vec{s}^{\mathsf{T}}_{tot}\cdot T[\hat{M}_{\bf{q}}](\vec{s}_{tot})=\langle T[\hat{M}_{\bf{q}}]\rangle_{\vec{s}_{tot}}=\frac{1}{8^{n}}(1+\frac{1}{4^{n}}|\braket{\bf{q}}{\psi_{tot}}|^2),\quad {\bf{q}}\in\mathbb{B}^{n} \quad . 
\end{equation}
To give a specific example, consider that at the end of the quantum evolution, we are interested in finding the probability that the qubit system is in  the state $\ket{000...0}$ after a measurement in that basis. This probability is $\vert \braket{000...0}{\psi_{tot}} \vert^2$, and can be calculated directly from the components of the simplex vector $ \vec{s}_{tot}$:
\begin{eqnarray}
\vert \braket{000...0}{\psi_{tot}} \vert^2 = (1-8^n s_{tot,1})^2 +(1-8^n s_{tot,5})^2  \quad .
\end{eqnarray}

Furthermore, we note that in a quantum system, the outcomes of measurements do not depend on the absolute phase of the wavefunction. As we mentioned above, different absolute phases for the quantum wavefunction result in different maps in the simplex. A critical point of consideration is whether the measurements we make in the simplex are independent of the phase ordering we choose for a particular state or not? It can in fact be proven that the phase ordering operations do not affect the measurements we make in the logical basis (see Appendix E for the proof) establishing the consistency of the phase ordering operations that we introduce in Appendix D.

For measurements more general than just a projection operator, a connection that is identical to the single qubit case presented in Eq.~(33) holds. That is, for any given quantum observable $\hat{A}$, quantum state $\ket{\psi_{tot}}$ and the corresponding simplex state $\vec{s}_{tot}=\varphi_{n}\ket{\psi_{tot}}$ (note here that the phase order has not been specified because of its irrelevance), we have the following connection between the quantum measurement and the simplex measurement (see Appendix E for the proof):
\begin{equation}
    \vec{s}_{tot}^{\mathsf{T}}\cdot\,T[\hat{A}](\vec{s}_{tot})=\langle T[\hat{A}]\rangle_{\vec{s}_{tot}}=\frac{1}{8^{n}}(1+\frac{1}{4^{n}}\langle\hat{A}\rangle_{\ket{\psi_{tot}}}) \quad . 
\end{equation}

\section{V: The Deutsch-Jozsa Algorithm}
In the Deutsch - Jozsa problem \cite{deutsch}, we are given a black box quantum computer known as an oracle that implements some function $f$.  The function $f$ takes $n$-bit binary values as input and produces either a 0 or a 1 as output for each such value. We are promised that the function is either constant (0 on all inputs or 1 on all inputs) or balanced (0 for exactly half of the input domain and 1 for the other half). The task then is to determine if $f$ is constant or balanced by using the oracle. For a classical deterministic algorithm, an exponential number of evaluations of the function are required. For a quantum algorithm, only a single quarry to the function $f$ is sufficient. The Deutsch-Jozsa algorithm is critically important in the history of quantum computation, since it was the first algorithm to explicitly show that there can be an exponential speed-up if quantum computing is used. 

One thing to note is that in the Deutsch - Jozsa quantum algorithm, complex values for the coefficients are not needed; i.e., only the real values and the signs are important. Because of that, for the mapping of each qubit, we need the components of the simplex vectors that store only the real parts of the coefficients; i.e., for each mapped qubit only $\vec{p}_0$ and $\vec{p}_1$ vectors introduced above in Section~III are sufficient.  In our implementation, we follow quite closely the main steps in the quantum algorithm. We start with $n+1$ simplex vectors. The first $n$ simplex vectors are initialized to their $\vec{p}_{0}$ state, while the final vector is in $\vec{p}_1$ state. Our initial state is therefore:
\begin{eqnarray}
  \frac{1}{8^{n+1}}  \left[\vec{u}^{\otimes_{n+1}}  +  \left(\overbrace{ \vec{p}_0 \otimes \vec{p}_0 \otimes ... \otimes \vec{p}_0}^{n\text{ simplex vectors}} \otimes \vec{p}_1 \right)  \right] \quad . 
\end{eqnarray}
We then proceed with applying a Hadamard gate to each of the simplex vectors to obtain:
\begin{eqnarray}
& & \frac{1}{8^{n+1}}  \left[ \vec{u}^{\otimes_{n+1}}  + \frac{1}{\sqrt{2^{n+1}}}  \left( \vec{p}_0 +\vec{p}_1 \right)  \otimes \left( \vec{p}_0 + \vec{p}_1 \right) \otimes ... \otimes \left(  \vec{p}_0 + \vec{p}_1 \right)  \otimes \left( \vec{p}_0 - \vec{p}_1 \right)   \right] \quad , \nonumber \\
 & = & \frac{1}{8^{n+1}}  \left[ \vec{u}^{\otimes_{n+1}}  + \frac{1}{\sqrt{2^{n+1}}}  \sum_{z=0}^{2^{n}-1} \vec{p}_{\bf{z}} \otimes \left( \vec{p}_0 - \vec{p}_1 \right)   \right] \quad .
\end{eqnarray}
\begin{figure}
    \centering
    \includegraphics[width=0.5\textwidth]{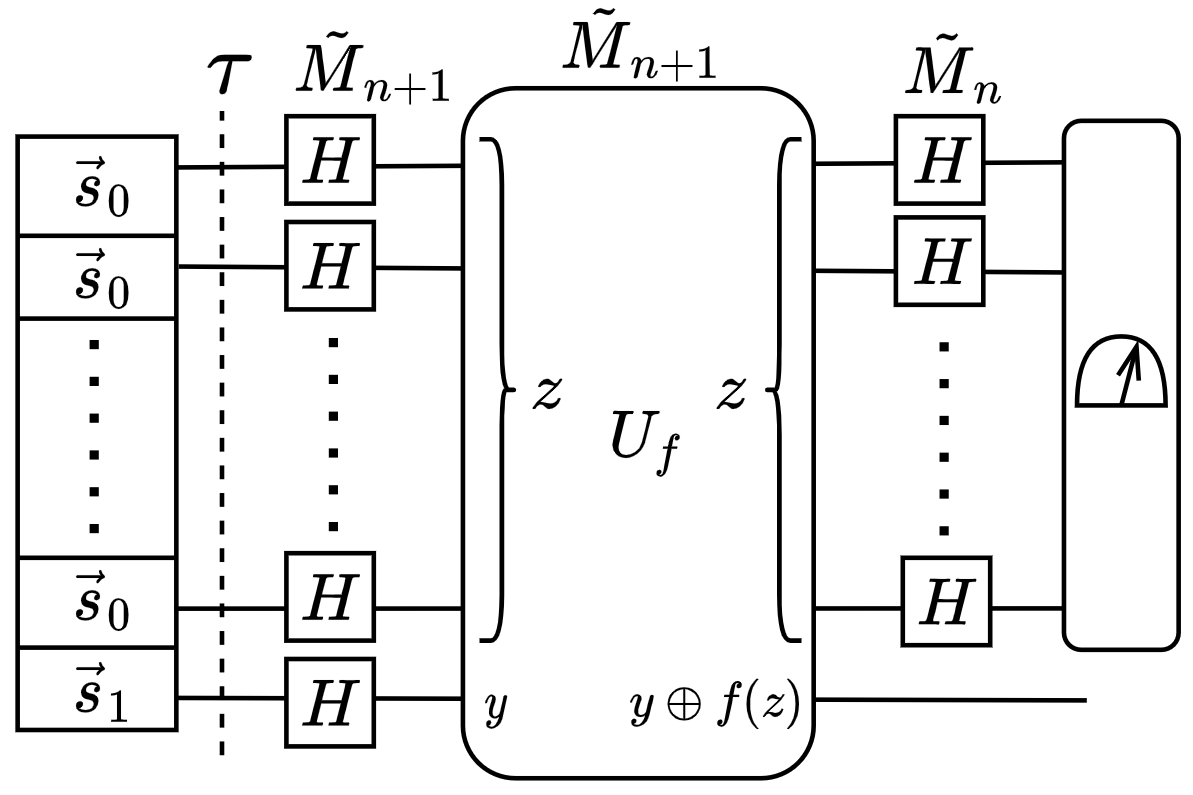}
    \caption{Simplex version of the Deutsch-Jozsa algorithm. The dashed line marked as $\tau$ indicates recursive application of the bi-affine map $\tau$ as shown in Eq.~(62). Therefore, the state before the application of Hadamard gates is as indicated in Eq.~(72). The quantum oracle evaluating the function $f$ can be represented by a unitary evolution, $\hat{U}_f$. This unitary evolution is encoded in the black box simplex transformation, $\tilde{M}_{n+1}[\hat{U}_f]$ such that it has the action as desired on the simplex states.}
\end{figure}
Now, similar to the quantum case, we have the function $f$ implemented as an oracle. The oracle maps the state $\vec{p}_{\bf{z}} \otimes \vec{p}_{y}$ to $\vec{p}_{\bf{z}} \otimes \vec{p}_{y'}$ where $y' = y \oplus f(z)$ and  $z$ denotes the decimal equivalent of $({\bf{z}})_2$. Here $\oplus$ denotes addition modulo 2. Applying this oracle to state above gives:
\begin{eqnarray}
 \frac{1}{8^{n+1}}  \left[ \vec{u}^{\otimes_{n+1}}  + \frac{1}{\sqrt{2^{n+1}}}  \sum_{z=0}^{2^{n}-1} \vec{p}_{\bf{z}} \otimes \left( \vec{p}_{0'} - \vec{p}_{1'} \right)   \right] \quad .
\end{eqnarray}
where we have $0'= 0 \oplus f(z)$ and $1' = 1  \oplus f(z) $ for the $(n+1)^{\text{th}}$ bit, respectively.  Noting that for each $z$, there are only two possibilities for $f(z)$, either 0 or 1. As a result, the above state actually equals to:
\begin{eqnarray}
 & = & \frac{1}{8^{n+1}}  \left[ \vec{u}^{\otimes_{n+1}}  + \frac{1}{\sqrt{2^{n+1}}}  \sum_{z=0}^{2^{n}-1} (-1)^{f(z)} \vec{p}_{\bf{z}} \otimes \left( \vec{p}_0 - \vec{p}_1 \right)   \right] \quad .
\end{eqnarray}
At this point, the $(n+1)^{\text{th}}$ vector is redundant and can be ignored. Focusing on just the first $n$ simplex vectors, we have the state:
\begin{eqnarray}
 \frac{1}{8^{n}}  \left[ \vec{u}^{\otimes_{n}}  + \frac{1}{\sqrt{2^{n}}}  \sum_{z=0}^{2^{n}-1} (-1)^{f(z)} \vec{p}_{\bf{z}}  \right] \quad .
\end{eqnarray}
Next, we apply the following $n$-bit Hadamard transform to the $\vec{p}_{\bf{z}}$ vector:
\begin{eqnarray}
\tilde{M}_n[\hat{H}^{\otimes_n}]\cdot\vec{p}_{\bf{z}} = \frac{1}{\sqrt{2^n}} \sum_{k=0}^{2^{n}-1} (-1)^{\bf{z}\cdot\bf{k}} \vec{p}_{\bf{k}} \quad .  
\end{eqnarray}
Here, the quantity ${\bf{z} \cdot \bf{k}} = z_1\cdot k_1 \oplus z_2 \cdot k_2 \oplus z_3 \cdot k_3 \oplus ... \oplus z_n \cdot k_n$ is the sum of the bitwise product and as above $\oplus$ denotes addition modulo 2.\\
Applying the Hadamard transformation of Eq.~(77) to the state of Eq.~(76), we then get:
\begin{eqnarray}
 &  & \frac{1}{8^{n}}  \left[ \vec{u}^{\otimes_{n}}  + \frac{1}{\sqrt{2^{n}}}  \sum_{z=0}^{2^{n}-1} (-1)^{f(z)} \tilde{M}_n[\hat{H}^{\otimes_n}]\cdot\vec{p}_{\bf{z}}  \right] \quad , \nonumber \\
 & = &  \frac{1}{8^{n}}  \left[ \vec{u}^{\otimes_{n}} + \frac{1}{\sqrt{2^{n}}}  \sum_{z=0}^{2^{n}-1} (-1)^{f(z)} \left(   \frac{1}{\sqrt{2^n}} \sum_{k=0}^{2^{n}-1} (-1)^{\bf{z} \cdot \bf{k}} \vec{p}_{\bf{k}}  \right)  \right] \quad , \nonumber \\
 & = &  \frac{1}{8^{n}}  \left[ \vec{u}^{\otimes_{n}} + \frac{1}{2^{n}}  \sum_{k=0}^{2^{n}-1} \sum_{z=0}^{2^{n}-1}  (-1)^{f(z)}  (-1)^{\bf{z} \cdot \bf{k}} \vec{p}_{\bf{k}}   \right] \quad . 
\end{eqnarray}
Now, the basic idea is that if the function $f$ is constant, then the sum $ \sum_{z=0}^{2^{n}-1}  (-1)^{f(z)}  (-1)^{\bf{z} \cdot \bf{k}}$ is $2^n$ only for $k=0$ and zero otherwise. The result which is identical to the quantum case is now stored in the corresponding $\vec{p}$ vector of the final probability distribution; that is if the function $f$ is constant, the final simplex vector has $\vec{p}_{\bf{k}}$ contribution for only $k=0$. This is achieved by a single operation of the oracle function $\tilde{M}_{n+1}[\hat{U}_f]$ to the overall simplex vector.

\section{VI: Quantum Fourier Transform}
Quantum Fourier transform is a method to achieve discrete Fourier transform in an exponentially large Hilbert space using only polynomial number of operations \cite{coppersmith}. In this section, we will first review the main steps in the Quantum Fourier Transform operation and then discuss its implementation in the probability space. Let's consider an exponentially large sequence of numbers, $x_j$, of length $L$: $\{0\leqslant x_j\leqslant 1:j\in\{0,1,\dots,L-1\}\}$. The discrete Fourier transform of such a sequence is given by the following expression: 
\begin{equation}
    y_{k}=\frac{1}{\sqrt{L}}\sum_{j=0}^{L-1}e^{2\pi i\,jk/L} x_{j}, \forall k\in\{0,1,\dots,L-1\}\quad .
\end{equation}
Because this transformation is unitary we can envision a quantum procedure that achieves the above transformation for the expansion coefficients of the quantum state in a certain basis. Quantum Fourier Transform, which forms a critical step in Shor's factoring algorithm, specifies a method for transformation of the components of the basis states in a manner identical to Eq.~(79).  More explicitly, for a sequence of length $L=2^{n}$, we design a quantum unitary evolution matrix $\hat{Q}_{n}$ on $n=\log_2L$ qubits which has the following action on a given logical basis state $\ket{j_1,j_2,\dots,j_n}\equiv \ket{j}$, 
\begin{equation}
    \hat{Q}_n\ket{j_1,j_2,\dots,j_n}=\hat{Q}_n\ket{j}\equiv\frac{1}{\sqrt{2^n}}\sum_{k=0}^{2^{n}-1}e^{2\pi i\, jk/2^{n}}\ket{k} \quad .
\end{equation}
where each $j_\nu,k_\nu\in\mathbb{B}$ and $j$, $k$ are the decimal equivalents of the binary representations: $(j_1\,j_2\,\cdots\,j_n)_2$ and $(k_1\,k_2\,\cdots\,k_n)_2$,  respectively. Following the definition of above, if we have a general state, 
\begin{equation}
    \ket{x}=\sum_{j=0}^{2^{n}-1}x_j\ket{j} 
\end{equation}
that stores the sequence $\{0\leqslant x_j\leqslant 1:j\in\{0,1,\dots,2^{n}-1\}\}$, then the application of the unitary operator $\hat{Q}_n$ to this state provides us with a state that stores the discrete Fourier transform of the aforementioned sequence of coefficients. This can be seen by noting that:
\begin{align}
    \hat{Q}_n\ket{x}&=\sum_{j=0}^{2^{n-1}}x_j\hat{Q}_n\ket{j}=\sum_{j=0}^{2^{n}-1}x_j\frac{1}{\sqrt{2^n}}\sum_{k=0}^{2^{n}-1}e^{2\pi i\,jk/2^{n}}\ket{k}\\
    &=\sum_{k=0}^{2^{n}-1}\overbrace{\frac{1}{\sqrt{2^n}}\left(\sum_{j=0}^{2^{n}-1}e^{2\pi i\,jk/2^{n}}x_j\right)}^{y_k}\ket{k}=\sum_{k=0}^{2^{n}-1}y_k\ket{k}=\ket{y},
\end{align}
While the unitary operator, $\hat{Q}_n$ is of dimension $2^n \times 2^n$ and acts on an exponentially large state space, remarkably, the Quantum Fourier Transform operation can be implemented using $O(n^2)$ single-qubit and two-qubit gates. This can most readily be seen by writing the effect of $\hat{Q}_n$ on a basis state $\ket{j}$ in the following product form: 
\begin{equation}
    \hat{Q}_n\ket{j}=\frac{1}{(\sqrt{2})^{n}}(\ket{0}+e^{2\pi i(0.j_n)_2}\ket{1})(\ket{0}+e^{2\pi i(0.j_{n-1}j_{n})_2}\ket{1})\cdots(\ket{0}+e^{2\pi i(0.j_{1}j_{2}j_{3}\dots j_n)_2}\ket{1}) \quad .
\end{equation} 
\begin{figure}
    \centering
    \includegraphics[width=0.8\textwidth]{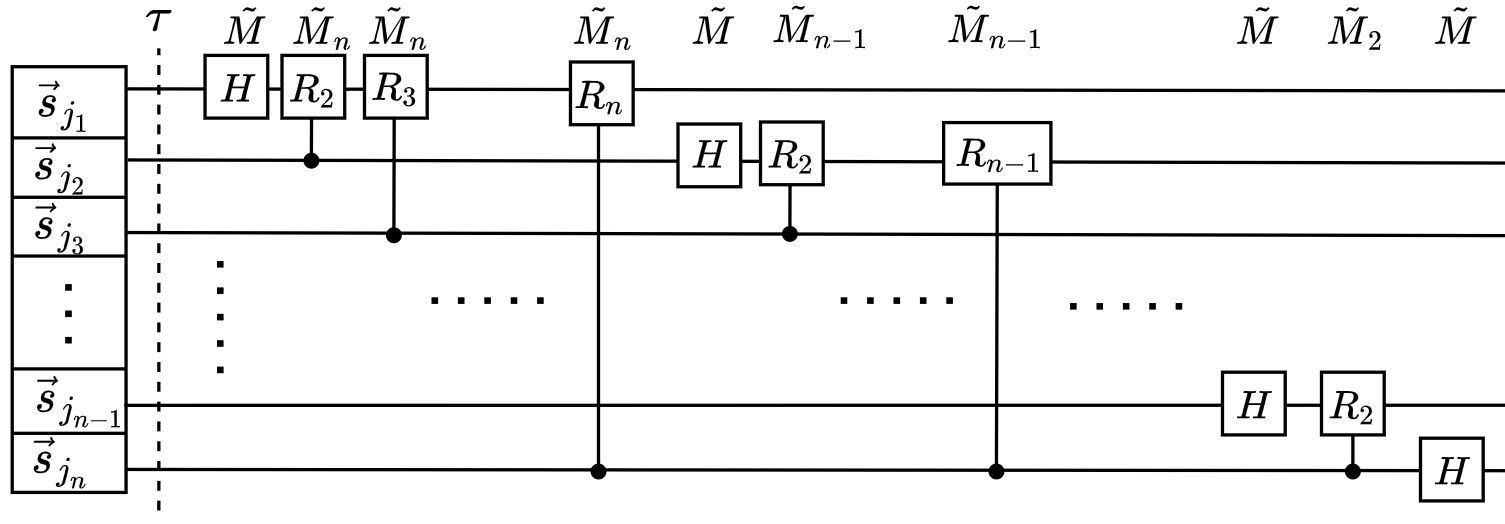}
    \caption{Circuit for implementing $\tilde{M}[\hat{Q}'_n]$. The dashed line marked as $\tau$ indicates recursive application of the bi-affine map $\tau$ as shown in Eq.~(62). The expression for the rotation gates $\hat{R}_k$ are provided in Eq.~(86). At the end of this circuit, identical to the quantum case, the output bits are in reverse order. Furthermore, there is also another complication, due to how the absolute phase of the wavefunction can be distributed (ordered) into individual simplex vectors.  To obtain the Fourier Transform operation identical to the quantum case, post processing involves application of a phase ordering operations followed by $\lfloor n/2\rfloor$ SWAP operations (see Fig.~7).}
\end{figure}
In order to implement the simplex version of the Quantum Fourier Transform, we first write the above product form in the probability space, in the tensor product of the $\vec{p}$ vectors:
\begin{equation}
    \tilde{M}[\hat{Q}_n]\cdot\vec{p}_{j_1,j_2\dots,j_n}=\frac{1}{(\sqrt{2})^n}(\vec{p}_0+\vec{P}_1(e^{2\pi i(0.j_n)_2}))\otimes(\vec{p}_0+\vec{P}_1(e^{2\pi i(0.j_{n-1}j_{n})_2}))\otimes\cdots\otimes(\vec{p}_0+\vec{P}_1(e^{2\pi i(0.j_{1}j_{2}j_{3}\dots j_n)_2})) \quad .
\end{equation}
There is a well-known procedure for implementing the Quantum Fourier Transform using a sequence of Hadamard gates and controlled-phase rotations. This procedure is first implemented in a reverse order on the qubits, and then transformed into the desired form by using a final set of SWAP operations. To implement the simplex analog of the Quantum Fourier Transform, we follow this procedure gate by gate.  We give a circuit diagram in Fig.~6 that generates the state in Eq.~(84) but in reverse order, we call this operation $\tilde{M}[\hat{Q}'_n]$. In Fig.~6, each gate $H$ is a Hadamard gate on that specific simplex vector, and  each controlled-rotation gate $\hat{R}_k$ is the following matrix: 
\begin{equation}
    \hat{R}_{k}\equiv\left(\begin{array}{cc}
				1& 0\\
				0& e^{2\pi i/2^k}
			\end{array}\right)\quad .
\end{equation}
We now outline the procedure step-by-step as follows. First, we start with the initial product state of the simplex vectors $\vec{s}_{j_1}\otimes\vec{s}_{j_2}\otimes\cdots\otimes\vec{s}_{j_n}$. This product state is obtained with the identical procedure that we discussed above, for example, in the Deutsch-Jozsa algorithm. Each qubit is mapped to a simplex vector using the mapping of Eq.~(3), $\ket{\psi_{j_i}} \rightarrow \vec{s}_{j_i}$. We then apply the recursive bi-affine map $\tau$ to get the quantum analogue of the initial product state $\ket{j}$ (cf. Eqs.~(61) and (62)): 
\begin{equation}
    \varphi_{n}\ket{j}=\frac{1}{8^{n}}(\vec{u}^{\otimes_n}+\vec{p}_{j_1}\otimes\vec{p}_{j_2}\otimes\cdots\otimes\vec{p}_{j_n})=\frac{1}{8^{n}}(\vec{u}^{\otimes_n}+\vec{p}_{j_1,j_2,\dots,j_n}) \quad .
\end{equation}
We then transform the first simplex vector, $\vec{s}_{j_1}$ by a sequence of $n-1$ controlled rotation gates followed by a Hadamard gate \footnote{For binary representations we use the right shift property: $2^{-\ell}(b_1\,b_2\,\cdots\,b_n\,.\,00\cdots)_2=(b_1\,b_2\,\cdots\,b_\ell\,.\,b_{\ell+1}\,\cdots\,b_n)_2$}. If we track this evolution gate-by-gate, below are the simplex states that are produced:
\begin{itemize}
    \item The first Hadamard gate  
    \begin{align*}
    (\tilde{M}[\hat{H}]\cdot\vec{p}_{j_1})\otimes\vec{p}_{j_2,j_3,\dots,j_n}&=\frac{1}{\sqrt{2}}(\vec{p}_0+(-1)^{j_1}\vec{p}_{1})\otimes(\vec{p}_{j_2,j_3,\dots,j_n})\\
    &=\frac{1}{\sqrt{2}}(\vec{p}_0+\vec{P}_{1}(e^{i\pi j_1}))\otimes\vec{p}_{j_2,j_3\dots,j_n}\\
    &=\frac{1}{\sqrt{2}}(\vec{p}_0+\vec{P}_{1}(e^{2\pi i(0.j_1)_2}))\otimes\vec{p}_{j_2,j_3\dots,j_n}
    \end{align*} 
    \item Controlled $\hat{R}_2$ rotation $\hat{C}^{(2,1)}_{\hat{R}_2}$: 
    \begin{align*}
    &\tilde{M}_n[\hat{C}^{(2,1)}_{\hat{R}_2}]\cdot\left(\frac{1}{\sqrt{2}}(\vec{p}_0+\vec{P}_{1}(e^{2\pi i(0.j_1)_2}))\otimes\vec{p}_{j_2,j_3\dots,j_n}\right)\\
    &=\frac{1}{\sqrt{2}}(\tilde{M}[\hat{R}_{2}^{j_2}]\cdot(\vec{p}_0+\vec{P}_{1}(e^{2\pi i(0.j_1)_2})))\otimes\vec{p}_{j_2,j_3,\dots,j_n}\\
    &=\frac{1}{\sqrt{2}}(\vec{p}_0+\vec{P}_{1}(e^{2\pi i((0.j_1)_2+j_2/2^2)}))\otimes\vec{p}_{j_2,j_3,\dots,j_n}=\frac{1}{\sqrt{2}}(\vec{p}_0+\vec{P}_{1}(e^{2\pi i(0.j_1j_2)_2}))\otimes\vec{p}_{j_2,j_3,\dots,j_n}
    \end{align*} 
    \item[{$\vdots$}]
    \item[{$\vdots$}]
    \item By continuing this sequence, after controlled $\hat{R}_n$ rotation $\hat{C}^{(n,1)}_{\hat{R}_n}$ the state will be: 
    \begin{align*}
    &\tilde{M}_n[\hat{C}^{(n,1)}_{\hat{R}_n}]\cdot\left(\frac{1}{\sqrt{2}}(\vec{p}_0+\vec{P}_{1}(e^{2\pi i(0.j_1j_2\dots j_{n-1})_2})))\otimes\vec{p}_{j_2,j_3\dots,j_n}\right)\\
    &=\frac{1}{\sqrt{2}}(\vec{p}_0+\vec{P}_{1}(e^{2\pi i(0.j_1j_2\dots j_{n})_2}))\otimes\vec{p}_{j_2,j_3,\dots,j_n}
    \end{align*} 
\end{itemize}
Using a procedure similar to above, we next transform the second simplex state, $\vec{s}_{j_2}$, via $n-2$ controlled rotations again starting with a Hadamard gate. For this case, the second simplex state will be transformed appropriately, producing the following output overall $\vec{p}$ vector:
\begin{equation}
    \frac{1}{(\sqrt{2})^2}(\vec{p}_0+\vec{P}_{1}(e^{2\pi i(0.j_1j_2\dots j_{n})_2}))\otimes(\vec{p}_0+\vec{P}_{1}(e^{2\pi i(0.j_2j_3\dots j_n)}))\otimes\vec{p}_{j_3,\dots,j_n}\quad .
\end{equation}
\begin{figure}
    \centering
    \includegraphics[width=0.8\textwidth]{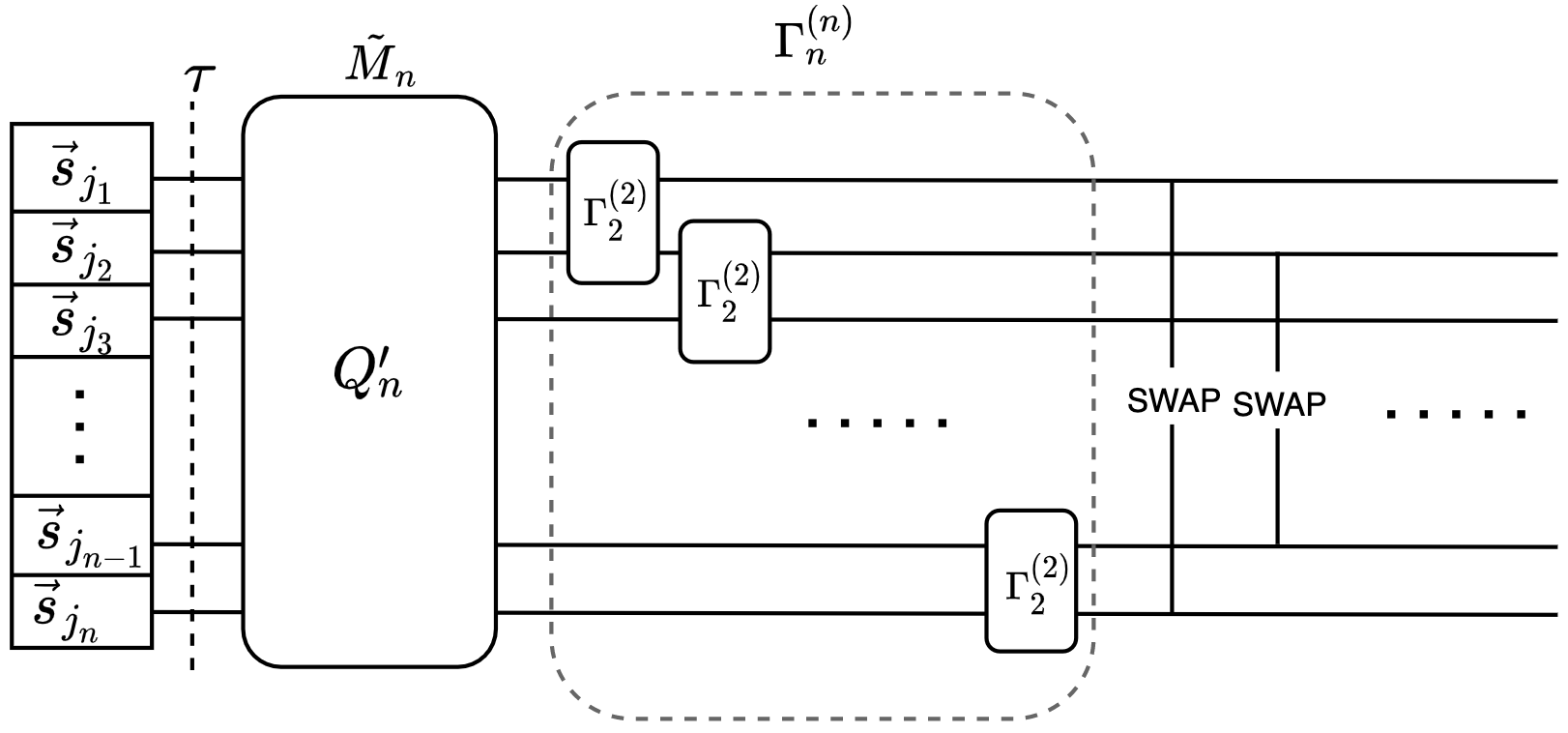}
    \caption{Post processing after the method $\tilde{M}[\hat{Q}'_n]$ proposed in Fig.~6. Here we show an example of making a first phase ordered state. The SWAP operation after the phase ordering operation $\Gamma^{(n)}_{n}$ will make up a first phase ordered state.} 
\end{figure}

As shown in Fig. 6, this procedure is continued until the final simplex state. At the end of the procedure, the last simplex vector, $\vec{s}_{j_n}$ is evolved by the application of a single Hadamard gate. The final output $\vec{p}$ vector can be rewritten in the same manner as before giving us the form required in Eq.~(84) but in reverse order,
\begin{equation}
    \frac{1}{(\sqrt{2})^n}(\vec{p}_0+\vec{P}_1(e^{2\pi i(0.j_{1}j_{2}j_{3}\dots j_n)_2}))\otimes\cdots\otimes(\vec{p}_0+\vec{P}_1(e^{2\pi i(0.j_{n-1}j_{n})_2}))\otimes(\vec{p}_0+\vec{P}_1(e^{2\pi i(0.j_n)_2}))\quad .
\end{equation}
This form achieves the quantum Fourier transform operation of Eq.~(84), but in reverse order. By applying  $\lfloor n/2\rfloor$ SWAP operation, we would accomplish the quantum Fourier transform operation of Eq.~(84). We next discuss the complication of the overall absolute phase of the wavefunction has on the simplex vectors. For the case of the quantum algorithm, because the phases are treated as numerical factors, the expressions in Eqs.~(80) and (84) are equivalent. However the same cannot be said for the corresponding simplex expressions. Therefore, for post processing we first apply a particular phase ordering operation and then $\lfloor n/2\rfloor$ SWAP operations. In Fig.~7, we give an example of constructing a specific phase order for the Fourier transformed states. After application of $\tilde{M}[\hat{Q}'_n]$, this specific phase ordering operation adds all the phases and stores it in the last state of all the product states when expanding the state in Eq.~(80). Then the linear $\lfloor n/2\rfloor$ SWAP operations will reverse each of the product states in the expansion  thereby shifting the position of the phases. The final state that we produce will be of the following form, which we refer to as the first phase-ordered state: 
\begin{equation}
    \frac{1}{(\sqrt{2})^n}\sum_{k_1\in\mathbb{B}}\cdots\sum_{k_n\in\mathbb{B}}\vec{P}^{(1)}_{k_1,\dots,k_n}(e^{2\pi i\,jk/2^{n}})=\frac{1}{(\sqrt{2})^n}\sum_{k_1\in\mathbb{B}}\cdots\sum_{k_n\in\mathbb{B}}\vec{P}_{k_1}(e^{2\pi i\,jk/2^{n}})\otimes\vec{p}_{k_2,\dots,k_n} \quad . 
\end{equation}
From this we can generate other phase ordered states by just applying appropriate order switching operations. The details of these operations are discussed in Appendix D. Since, we have the capacity to order phases of the states after performing the Fourier transform operation, it is important that we explicitly mention this in the operator notation. We do this by using an additional order subscript $\sigma$: $\tilde{M}[\hat{Q}_n]_\sigma$, where $\sigma\in\{1,2,\dots,n\}$ denotes the final phase ordering of the state after the Fourier transform operation has been performed. The transform $T$ over the simplex for the Fourier transform operation $\tilde{M}[\hat{Q}_n]_\sigma$ will be denoted by $T[\hat{Q}^{(n)}_\sigma]$.

Finally, we note that $T[\hat{Q}^{(n)}_\sigma]$ has the desired effect on a general state as well. As before starting with a general state in some phase order $\sigma$ \footnote{ For a given binary representation ${\bf{q}}$ its decimal equivalent $q$ is represented by: $[{\bf{q}}]=q$.}, 
\begin{equation}
    \varphi^{\sigma}_n\ket{x}=\vec{s}^{(\sigma)}(x)\doteq\frac{1}{8^n}(\vec{u}^{\otimes_n}+\sum_{q=0}^{2^{n}-1}\vec{P}^{(\sigma)}_{\bf{q}}(x_q)),\quad {\bf{q}}\in\mathbb{B}^n, q=[{\bf{q}}],
\end{equation}
we apply the Fourier transform operation and simplify, 
\begin{align}
    T[Q^{(n)}_{\sigma}](\vec{s}^{(\sigma)}(x))&=\frac{1}{8^{n}}\left(\vec{u}^{\otimes_n}+\tilde{M}[\hat{Q}_n]_\sigma\cdot\left(\sum_{q=0}^{2^{n}-1}\vec{P}^{(\sigma)}_{\bf{q}}(x_q)\right)\right)\\
    &=\frac{1}{8^{n}}\left(\vec{u}^{\otimes_n}+\left(\sum_{q=0}^{2^{n}-1}\tilde{M}[\hat{Q}_n]_\sigma\cdot\vec{P}^{(\sigma)}_{\bf{q}}(x_q)\right)\right)\\
    &=\frac{1}{8^{n}}\left(\vec{u}^{\otimes_n}+\sum_{q=0}^{2^{n}-1}\frac{1}{(\sqrt{2})^n}\sum_{k=0}^{2^n-1}\vec{P}^{(\sigma)}_{\bf{k}}(e^{2\pi i\,qk/2^{n}}x_q)\right),\quad k=[\bf{k}]\\
    &=\frac{1}{8^{n}}\left(\vec{u}^{\otimes_n}+\sum_{k=0}^{2^n-1}\vec{P}^{(\sigma)}_{\bf{k}}\left(\overbrace{\frac{1}{(\sqrt{2})^n}\sum_{q=0}^{2^{n}-1}e^{2\pi i\,qk/2^{n}}x_q}^{y_k}\right)\right)\\
    &=\frac{1}{8^{n}}\left(\vec{u}^{\otimes_n}+\sum_{k=0}^{2^n-1}\vec{P}^{(\sigma)}_{\bf{k}}(y_k)\right)=\vec{s}^{(\sigma)}(y)=\varphi^{\sigma}_{n}\ket{y} \quad . 
\end{align}
We, therefore, see that the final state indeed stores the Fourier transform of the input sequence in the same phase order. Note that to go from Eq.~(94) to (95) we have used the additive property from Eq.~(12).

\section{VII: Conclusions and Future Directions}

In conclusion, we have discussed a new approach to simulate quantum algorithms using classical probabilistic bits and circuits. Each qubit (a two-level quantum system) is initially mapped to a vector in an eight dimensional probability space. Due to the identical tensor product structure of combining multiple quantum systems as well as multiple probability spaces, $n$ qubits are then mapped to a tensor product of $n$ 8-dimensional probabilistic vectors (i.e., the Hilbert space of dimension $2^n$ is mapped to a probability space of dimension $8^n$). After this initial mapping, we showed how to implement analogs of single-qubit and two-qubit gates in the probability simplex.  Remarkably, these results show that an exponentially large number of complex coefficients in the quantum evolution can be tracked in the probability space with a similar number of operations performed in the probability simplex. We also discussed how to simulate (1) the Deutsch-Jozsa algorithm in the probability space, and (2) the Quantum Fourier transform in the probability space. Identical to the quantum case, implementing the Quantum Fourier Transform in the probability space requires a polynomial number of gates (in an exponentially large probability space). 

 Our work shows that the initial state and the evolution of an $n$-qubit quantum computer can be captured using a polynomial number of fully-correlated classical random variables and affine circuits. One exciting future direction is to show if our approach constitutes a truly efficient simulation of quantum evolution and whether there is an exponential overhead hiding in a certain aspect of our formalism. As the state of the quantum computer evolves in the Hilbert space, the entries of the simplex vector (i.e., the specific joint probabilities) become exponentially small (similar to how the magnitudes of the complex coefficients become exponentially small in a quantum computer). As a result, we believe understanding how the evolution is affected by noise is critical. To be able to claim efficient simulation, a detailed study of noise and error correction in probabilistic circuits of the form that we describe here is essential. This is one clear future research direction. We think it is possible that error correction is more straightforward with classical random variables in the probability space, since we are allowed to make a polynomial number of copies of the individual simplex vectors, and introduce a redundancy.  

 We also note that the issue of noise and error correction is still an active research area for quantum computers. It is known that the threshold theorem is not applicable when there is correlated noise affecting all the qubits simultaneously in the quantum computer \cite{aharonov,terhal,aharonov2,preskill}. Recent work has shown that such errors can happen when the qubits are coupled to a common bosonic bath (which, inevitably happens in every quantum computer) \cite{mucciolo1,mucciolo2,hutter}. We discussed how correlated decay between the qubit levels causes an error on each qubit that scales with the number of qubits, thereby again violating one of the basic assumptions of the threshold theorem (the assumption that the gate errors can be assumed to be smaller than the a certain threshold) \cite{lemberger}.
 
We also note that, another important issue is that while the $\vec{p}$ vector tracks the complex coefficients of the quantum evolution, it  is the total simplex vector, $\vec{s}_{tot}$, that is physical and that contains the probabilities (and the final measurements are performed on this vector). In other words, the quantum evolution is tracked in the deviation of the probabilities from the uniform distribution, $\vec{u}$. As a result, the individual entries of the simplex vector, $\vec{s}_{tot}$, always remain exponentially small. This is different from the quantum case. In quantum systems, at the end of the evolution, the probability can be concentrated at a certain state. In our formalism, tracking this evolution, the entries of the $\vec{p}$ vector would concentrate at a certain state, but this would not happen for the $\vec{s}_{tot}$ vector (due to the initial uniform distribution in the definition of the $\vec{s}_{tot}$ vector). An important open question is if perform measurements on the final output $\vec{s}_{tot}$ vector (i.e. if we sample from the final joint distribution), how efficiently can we simulate the final measurement outcomes of the $n$-qubit quantum system (keeping in mind that we are allowed to make a polynomial number of copies of the simplex vector $\vec{s}_{tot}$?)

 We believe the approach presented here has practical and fundamental implications. On the practical side, as mentioned above, our approach may provide a unique way to simulate quantum systems that may be more efficient than currently possible. Within this context, an exciting immediate experimental direction is to experimentally demonstrate the simplex transformations for a single qubit that we have discussed. It may be possible to extend the recent experimental work of Datta and colleagues on probabilistic bits (p-bits) \cite{datta1,datta2}. One near-term future goal would be to observe the analog of the single-qubit Hadamard gate and ``Rabi rotations" in the simplex using an appropriate circuit acting on 3~$p$-bits. We have discussed a specific procedure for implementing ``Rabi rotations" using classical random variables and probabilities in our recent work \cite{deniz_rudhy}. Another goal would be to implement the analog two-qubits gates using correlation-inducing operations on multiple $p$-bits. On the fundamental side, we have shown that $n$-qubit state of a quantum computer can be tracked in the deviations of probabilities from a uniform distribution. For capturing the initial state and time evolution of a quantum system, the mathematical structure of complex wavefunctions that live in a Hilbert space is not necessary. We think it is also possible that progress along the above posed questions will help clarify the quantum/classical boundary \cite{niklas,mucciolo,pittenger}, as well as the quantum measurement problem \cite{zurek1,zurek2}. We also note that, throughout this paper, we have focused on simulating quantum algorithms using classical probabilistic bits and circuits. We have not gone into a detailed discussion of concepts of entanglement \cite{bell1,realq1,realq2}, nonlocality \cite{barrett2}, contextuality \cite{spekkens2}, or the reality of the quantum state \cite{montina,rudolph2,rudolph3}. A rigorous discussion of these important concepts is beyond the scope of this work.

We finally note that Quantum Fourier Transform is arguably the most important step in the celebrated Shor's factoring algorithm \cite{shor,ekert}. An exciting future direction is to extend our analysis and provide the specific procedures for implementing an analog Shor's factoring algorithm in the probability simplex.

\section{VIII. Acknowledgements}

We would like to thank Ben Lemberger and Volkan Rodoplu for many helpful discussions. D. D. Yavuz would also would like to thank Bin Yan for an early discussion on the subject. This work was supported by the National Science Foundation (NSF) Grant No. 2016136 for the QLCI center Hybrid Quantum Architectures and Networks (HQAN), and also by the University of Wisconsin-Madison, through the Vilas Associates award.

\newpage
\section{Appendix A. Rabi Rotations and Phase Gates in Simplex Space}
The unitary operation for a general Rabi rotation with angle $\theta$ is given by the following action on the logical basis states:
\begin{align}
    \hat{Y}_{\theta}\ket{0}&=\cos\left(\frac{\theta}{2}\right)\,\ket{0}+\sin\left(\frac{\theta}{2}\right)\,\ket{1} \quad , \\
    \hat{Y}_{\theta}\ket{1}&=\sin\left(\frac{\theta}{2}\right)\,\ket{0}-\cos\left(\frac{\theta}{2}\right) \ket{1} \quad . 
\end{align}
In this logical qubit basis, this operation can be represented by the following $2 \times 2$ unitary matrix:
\begin{equation}
    \hat{Y}_{\theta}=\left(\begin{array}{cc}
        \cos\left(\frac{\theta}{2}\right) & \sin\left(\frac{\theta}{2}\right) \\
        \sin\left(\frac{\theta}{2}\right) & -\cos\left(\frac{\theta}{2}\right)
    \end{array}\right)\quad .
\end{equation}
Similarly, the single qubit phase gate $\hat{Z}_{\phi}$ has the following action on the logical states:
\begin{equation}
    \hat{Z}_{\phi}\ket{0}=\ket{0} \quad , \quad \hat{Z}_{\phi}\ket{1}=e^{i\phi}\ket{1} \quad . 
\end{equation}
This phase gate, $\hat{Z}_{\phi}$, can be represented in the same matrix notation as:
\begin{equation}
    \hat{Z}_{\phi}=\left(\begin{array}{cc}
        1 & 0 \\
        0 & e^{i\phi}
    \end{array}\right)=\left(\begin{array}{cc}
        1 & 0 \\
        0 & \cos\phi
    \end{array}\right)+i\left(\begin{array}{cc}
        0 & 0 \\
        0 & \sin\phi
    \end{array}\right)\quad .
\end{equation}
\begin{figure}
    \centering
    \includegraphics[width=0.6\textwidth]{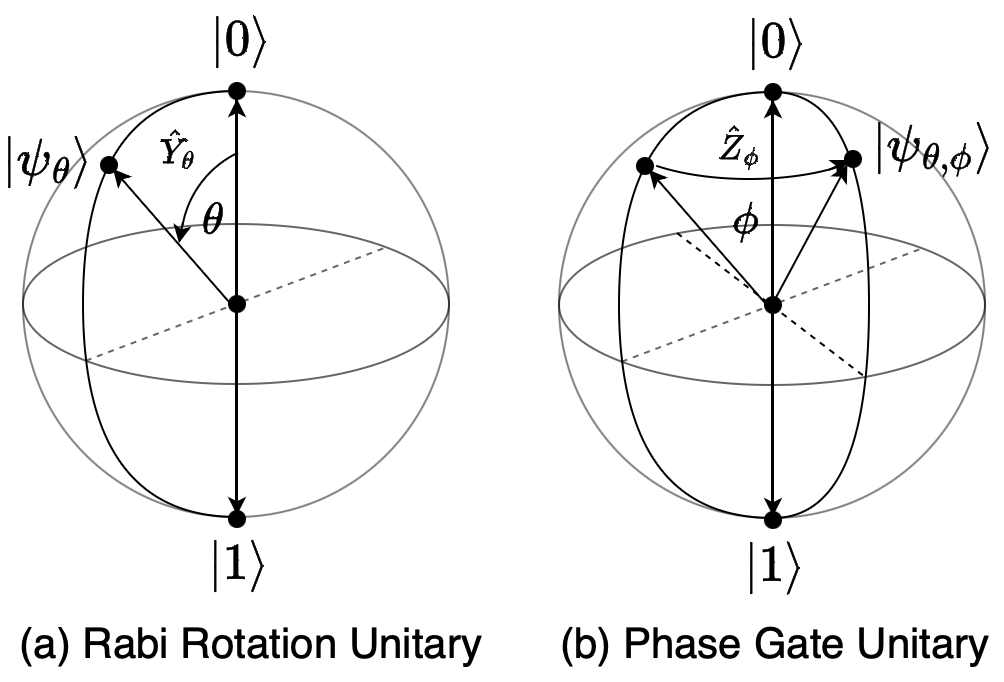}
    \caption{This figure illustrates the action of Rabi rotations $\hat{Y}_{\theta}$ and Phase gates $\hat{Z}_{\phi}$ on a qubit state in the Bloch sphere picture. We note that these two rotations are sufficient to steer the qubit on the Bloch sphere to any location $(\theta,\phi)$.}
\end{figure}
The actions of these operators in the Bloch sphere picture are shown visually in Fig. 8(a) and 8(b), respectively. Using the linear operator representations in Eq.~(99) and Eq.~(101), we can map these operations to their simplex counterparts respectively. The simplex transformation matrix for the Rabi rotations is given by (cf. Eq~(19)): 
\begin{equation}
    \tilde{M}[\hat{Y}_{\theta}]=\left( \begin{array}{cc|cc|cc|cc}
\cos\left(\frac{\theta}{2}\right) & \sin\left(\frac{\theta}{2}\right) & 0 & 0 &  0 & 0 & 0 & 0 \\
\sin\left(\frac{\theta}{2}\right) & -\cos\left(\frac{\theta}{2}\right) & 0 & 0 & 0 & 0 & 0 & 0 \\ 
\hline
0 & 0 & \cos\left(\frac{\theta}{2}\right) & \sin\left(\frac{\theta}{2}\right) & 0 & 0 & 0 & 0 \\
0 & 0 & \sin\left(\frac{\theta}{2}\right) & -\cos\left(\frac{\theta}{2}\right) & 0 & 0 & 0 & 0 \\
\hline
0 & 0 & 0 & 0 & \cos\left(\frac{\theta}{2}\right) & \sin\left(\frac{\theta}{2}\right) & 0 & 0 \\
0 & 0 & 0 & 0 & \sin\left(\frac{\theta}{2}\right) & -\cos\left(\frac{\theta}{2}\right) & 0 & 0 \\
\hline
0 & 0 & 0 & 0 & 0 & 0 & \cos\left(\frac{\theta}{2}\right) & \sin\left(\frac{\theta}{2}\right) \\
0 & 0 & 0 & 0 & 0 & 0 & \sin\left(\frac{\theta}{2}\right) & -\cos\left(\frac{\theta}{2}\right) \\
\end{array} \right) \quad .
\end{equation}
From $\tilde{M}[\hat{Y}_{\theta}]$ we can further define the affine transform $T[\hat{Y}_{\theta}]$ that can be applied to a simplex state to steer the state along the longitude of the Bloch sphere. Similarly, the transformation matrix for implementing the phase gate will be (again using Eq.~(19)):
\begin{equation}
    \tilde{M}[\hat{Z}_{\phi}]=\left( \begin{array}{cc|cc|cc|cc}
1 & 0 & 0 & 0 &  0 & 0 & 0 & 0 \\
0 & \cos\phi & 0 & 0 & 0 & 0 & 0 & \sin\phi \\ 
\hline
0 & 0 & 1 & 0 & 0 & 0 & 0 & 0 \\
0 & 0 & 0 & \cos\phi & 0 & \sin\phi & 0 & 0 \\
\hline
0 & 0 & 0 & 0 & 1 & 0 & 0 & 0 \\
0 & \sin\phi & 0 & 0 & 0 & \cos\phi & 0 & 0 \\
\hline
0 & 0 & 0 & 0 & 0 & 0 & 1 & 0 \\
0 & 0 & 0 & \sin\phi & 0 & 0 & 0 & \cos\phi \\
\end{array} \right) \quad .
\end{equation}
As before, this provides us with the affine transform $T[\hat{Z}_{\phi}]$ for steering states along the latitudes of the Bloch sphere. Using these two affine transformations ($T[\hat{Y}_{\theta}]$ and $T[\hat{Z}_{\phi}]$), we can steer the simplex state as the quantum wavefunction evolves to any corresponding point $(\theta,\phi)$ on the Bloch sphere.

\section{Appendix B. Proof of Measurement Correspondence}
The measurement correspondence that we aim to prove is: 
\begin{equation}
    \bra{\psi}\hat{A}\ket{\psi}=(\vec{p}^{\mathsf{T}}\cdot\tilde{M}[\hat{A}]\cdot\vec{p})/2\quad .
\end{equation}
To prove this statement we expand the left hand side and the right hand side into their real and imaginary components. Again writing the wavefunction $\ket{\psi}$ as: 
\begin{equation}
    \ket{\psi}\equiv\vec{x}+i\vec{y}
\end{equation}
where, $\vec{x}\equiv \Re\ket{\psi}$, $\vec{y}\equiv\Im\ket{\psi}$ and, $\hat{A}=\Re(\hat{A})+i\Im(\hat{A})$ the left hand side then can be expanded as follows: 
\begin{align}
    \nonumber \bra{\psi}\hat{A}\ket{\psi}&=(\vec{x}^{\mathsf{T}}-i\vec{y}^{\mathsf{T}})\cdot(\Re(\hat{A})+i\Im(\hat{A}))\cdot(\vec{x}+i\vec{y})\\
    \nonumber &=(\vec{x}^{\mathsf{T}}\cdot\Re(\hat{A})\cdot\vec{x}-\vec{x}^{\mathsf{T}}\cdot\Im(\hat{A})\cdot\vec{y}+\vec{y}^{\mathsf{T}}\cdot\Re(\hat{A})\cdot\vec{y}+\vec{y}^{\mathsf{T}}\cdot\Im(\hat{A})\cdot\vec{x})\\
    &+i(\vec{x}^{\mathsf{T}}\cdot\Re(\hat{A})\cdot\vec{y}+\vec{x}^{\mathsf{T}}\cdot\Im(\hat{A})\cdot\vec{x}-\vec{y}^{\mathsf{T}}\cdot\Re(\hat{A})\cdot\vec{x}+\vec{y}^{\mathsf{T}}\cdot\Im(\hat{A})\cdot\vec{y}) \quad .
\end{align}
If $\hat{A}$ is an observable then it is a Hermitian operator, implying that $\Re(\hat{A})$ is symmetric and $\Im(\hat{A})$ is anti-symmetric. As a result:
\begin{equation}
\vec{x}^{\mathsf{T}}\cdot\Im(\hat{A})\cdot\vec{x}=\vec{y}^{\mathsf{T}}\cdot\Im(\hat{A})\cdot\vec{y}=0
\end{equation}
and, 
\begin{equation}
    \vec{x}^{\mathsf{T}}\cdot\Re(\hat{A})\cdot\vec{y}=(\vec{x}^{\mathsf{T}}\cdot\Re(\hat{A})\cdot\vec{y})^{\mathsf{T}}=\vec{y}^{\mathsf{T}}\cdot\Re(\hat{A})\cdot\vec{x}\quad .
\end{equation}
Hence, the imaginary part of the right hand side of Eq.~(106) is identically equal to zero because of the Hermiticity of $\hat{A}$, 
 which reduces the left hand side of Eq.~(104) to:
\begin{equation}
    \bra{\psi}\hat{A}\ket{\psi}=\vec{x}^{\mathsf{T}}\cdot\Re(\hat{A})\cdot\vec{x}-\vec{x}^{\mathsf{T}}\cdot\Im(\hat{A})\cdot\vec{y}+\vec{y}^{\mathsf{T}}\cdot\Re(\hat{A})\cdot\vec{y}+\vec{y}^{\mathsf{T}}\cdot\Im(\hat{A})\cdot\vec{x}\quad .
\end{equation}
The right hand side of Eq.~(104) can be evaluated by explicitly writing  $\vec{p}$ and $\tilde{M}[\hat{A}]$ according to the definitions in Section III:
\begin{align}
    \nonumber \frac{1}{2}(\vec{p}^{\mathsf{T}}\cdot\tilde{M}[\hat{A}]\cdot\vec{p})&=\frac{1}{2}\left(\begin{array}{r r r r}\vec{x}^{\mathsf{T}}&-\vec{x}^{\mathsf{T}}&\vec{y}^{\mathsf{T}}&-\vec{y}^{\mathsf{T}}\end{array}\right)\left(\begin{array}{c| c| c| c}
		\Re(\hat{A})&O&O&\Im(\hat{A})\\
        \hline
		O&\Re(\hat{A})&\Im(\hat{A})&O\\
        \hline
		\Im(\hat{A})&O&\Re(\hat{A})&O\\
        \hline
		O&\Im(\hat{A})&O&\Re(\hat{A})
	\end{array}\right)\left(\begin{array}{r}
		\vec{x}\\
		-\vec{x}\\
		\vec{y}\\
		-\vec{y}
	\end{array}\right)\\
    \nonumber &=\frac{1}{2}\left(\begin{array}{r r r r}\vec{x}^{\mathsf{T}}&-\vec{x}^{\mathsf{T}}&\vec{y}^{\mathsf{T}}&-\vec{y}^{\mathsf{T}}\end{array}\right)\left(\begin{array}{r}
		\Re(\hat{A})\cdot\vec{x}-\Im(\hat{A})\cdot\vec{y}\\
		-(\Re(\hat{A})\cdot\vec{x}-\Im(\hat{A})\cdot\vec{y})\\
		\Im(\hat{A})\cdot\vec{x}+\Re(\hat{A})\cdot\vec{y}\\
		-(\Im(\hat{A})\cdot\vec{x}+\Re(\hat{A})\cdot\vec{y})
	\end{array}\right)\\
    &=\vec{x}^{\mathsf{T}}\cdot\Re(\hat{A})\cdot\vec{x}-\vec{x}^{\mathsf{T}}\cdot\Im(\hat{A})\cdot\vec{y}+\vec{y}^{\mathsf{T}}\cdot\Re(\hat{A})\cdot\vec{y}+\vec{y}^{\mathsf{T}}\cdot\Im(\hat{A})\cdot\vec{x} \quad.
\end{align}
This expression is exactly equal to the reduced left hand side in Eq.~(109) proving the original expression written in Eq.~(104).

\section{Appendix C. The simplex tensor operation}
We start with the definition of the bivalent simplex tensor operation $\sotimes$, which was defined above in Section~IV: 
\begin{equation}
    \vec{s}_{12}=\vec{s}_{1}\sotimes\vec{s}_2=\frac{1}{2}(\vec{s}_1\otimes\vec{s}_2+\Pi(\vec{s}_1)\otimes\Pi(\vec{s}_2))=\frac{1}{8^2}(\vec{u}^{\otimes_2}+\vec{p}_1\otimes\vec{p}_2) \quad . 
\end{equation}
As we discussed above, the main idea behind this operation is that it is an equal statistical mixture of $\vec{s}_1 \otimes \vec{s}_2$ and $\Pi(\vec{s}_1) \otimes  \Pi (\vec{s}_2)$, so that the above mentioned cross terms are eliminated producing a combined vector with the desired form, $\vec{s}_{12}$. We note that it is evident from above that this operation is closed for two vectors, since the final state vector $\vec{s}_{12}$ has the same form of deviations over uniform distribution. Next we must show that the operation $\sotimes$ remains closed even for more than two state vectors. We prove this by induction. We know that the operation is closed for $n=2$ and taking it to be true for $n=r-1$ we can show that it holds true for $n=r$. The form of the state vector for $n=r-1$ will be
\begin{equation}
    \vec{s}_{1,r-1}=\frac{1}{8^r}(\vec{u}^{\otimes_{r-1}}+\bigotimes_{k=1}^{r-1}\vec{p}_{k}) \quad . 
\end{equation}
We next combine the vector $\vec{s}_{1,r-1}$ with $\vec{s}_r$, by explicitly evaluating each of the two components:
\begin{equation}
    \vec{s}_{1,r-1}\otimes\vec{s}_r=\frac{1}{8^r}(\vec{u}^{\otimes_{r-1}}+\bigotimes_{k=1}^{r-1}\vec{p}_k)\otimes(\vec{u}+\vec{p}_{r})=\frac{1}{8^{r}}(\vec{u}^{\otimes_r}+\vec{u}^{\otimes_{r-1}}\otimes\vec{p}_r+\bigotimes_{k=1}^{r-1}\vec{p}_{k}\otimes\vec{u}+\bigotimes_{k=1}^{r}\vec{p}_{k})
\end{equation}
and,
\begin{equation}
    \Pi(\vec{s}_{1,r-1})\otimes\Pi(\vec{s}_r)=\frac{1}{8^r}(\vec{u}^{\otimes_{r-1}}-\bigotimes_{k=1}^{r-1}\vec{p}_k)\otimes(\vec{u}-\vec{p}_{r})=\frac{1}{8^{r}}(\vec{u}^{\otimes_r}-\vec{u}^{\otimes_{r-1}}\otimes\vec{p}_r-\bigotimes_{k=1}^{r-1}\vec{p}_{k}\otimes\vec{u}+\bigotimes_{k=1}^{r}\vec{p}_{k})
\end{equation}
We then use the definition of $\sotimes$ operation, which is an equal statistical mixture of the above evaluated two terms:
\begin{equation}
    \vec{s}_{1r}=\vec{s}_{1,r-1}\sotimes\vec{s}_r=\frac{1}{2}(\vec{s}_{1,r-1}\otimes\vec{s}_r+\Pi(\vec{s}_{1,r-1})\otimes\Pi(\vec{s}_r))=\frac{1}{8^r}(\vec{u}^{\otimes_{r}}+\bigotimes_{k=1}^{r}\vec{p}_{k}) \quad . 
\end{equation}
This completes the proof since the vector $\vec{s}_{1r}$ has the desired form. 
Now that we have established closure we next prove associativity. For three state vectors, the combinations can be made in any sequence such that we maintain the overall order of the state vectors, to show this we expand the two sequences: 
\begin{equation}
    (\vec{s}_{1}\sotimes\vec{s}_2)\sotimes\vec{s}_3=\vec{s}_{12}\sotimes\vec{s}_{3}=\frac{1}{8^3}(\vec{u}^{\otimes_3}+\vec{p}_1\otimes\vec{p}_2\otimes\vec{p}_3)
\end{equation}
this is can be seen by substituting $r=3$ in Eq.~(115). The other sequence is,
\begin{equation}
    \vec{s}_{1}\sotimes(\vec{s}_2\sotimes\vec{s}_3)=\vec{s}_{1}\sotimes\vec{s}_{23}=\frac{1}{2}(\vec{s}_1\otimes\vec{s}_{23}+\Pi(\vec{s}_1)\otimes\Pi(\vec{s}_{23}))
\end{equation}
where, 
\begin{equation}
    \vec{s}_1\otimes\vec{s}_{23}=\frac{1}{8^3}(\vec{u}+\vec{p}_1)\otimes(\vec{u}^{\otimes_2}+\vec{p}_2\otimes\vec{p}_3)=\frac{1}{8^3}(\vec{u}^{\otimes_3}+\vec{u}\otimes\vec{p}_2\otimes\vec{p}_3+\vec{p}_1\otimes\vec{u}^{\otimes_2}+\vec{p}_1\otimes\vec{p}_2\otimes\vec{p}_3)
\end{equation}
and, 
\begin{equation}
    \Pi(\vec{s}_1)\otimes\Pi(\vec{s}_{23})=\frac{1}{8^3}(\vec{u}-\vec{p}_1)\otimes(\vec{u}^{\otimes_2}-\vec{p}_2\otimes\vec{p}_3)=\frac{1}{8^3}(\vec{u}^{\otimes_3}-\vec{u}\otimes\vec{p}_2\otimes\vec{p}_3-\vec{p}_1\otimes\vec{u}^{\otimes_2}+\vec{p}_1\otimes\vec{p}_2\otimes\vec{p}_3)
\end{equation}
proving that the sequence of combination does not matter. Lastly, we end this Appendix by noting that the simplex tensor operation $\sotimes$ is inherently distributive as it derives its definition form a bi-affine map $\tau$ introduced in the main text Eq.~(38).

\section{Appendix D. Ordering of Phases}
We first utilize the notation introduced in the main manuscript to understand ordering of phases and then introduce the ordering operators next. 
In the most general case, note that a single qubit state $\ket{\psi}=c_0\ket{0}+c_1\ket{1}$
is mapped as follows,
\begin{equation}
    \varphi\ket{\psi}=\vec{s}(\psi)\doteq\frac{1}{8}(\vec{u}+\vec{P}_{0}(c_0)+\vec{P}_{1}(c_1))=\frac{1}{8}(\vec{u}+\sum_{b\in\mathbb{B}}\vec{P}_{b}(c_b)),
\end{equation}
where,
\begin{equation}
    \vec{P}_{b}(c)=\vec{\gamma}\otimes \Re(c\ket{b})+\vec{\gamma}'\otimes \Im(c\ket{b})=(\Re(c)\vec{\gamma}+\Im(c)\vec{\gamma}')\otimes\ket{b}
\end{equation}
such that $\vec{P}_{b}(1)=\vec{p}_{b}$, $\vec{P}_{b}(0)=\vec{0}$. 

We start by considering the mapping of two qubits in logical states $\ket{b}$ and $\ket{b'}$, respectively. Let's associate two absolute phases, $\phi$ and $\phi'$, with these logical states. When we consider mapping of this system to the probability space, using the above notation, we would have:
\begin{equation}
    re^{i\phi}\ket{b}\otimes r'e^{i\phi'}\ket{b'}=rr'e^{i(\phi+\phi')}\ket{bb'}\longrightarrow \vec{P}_{b}(re^{i\phi})\otimes\vec{P}_{b'}(r'e^{i\phi'})=rr'\vec{P}_{b}(e^{i\phi})\otimes\vec{P}_{b'}(e^{i\phi'}) \quad . 
\end{equation}
We note that,  although the phases that each state carry in the quantum case commute and can be associated to any state, the same for the corresponding states in the simplex version is not true. More explicitly, for the quantum case, all the below three expressions refer to exactly the same quantum state of the two-qubit wavefunction:
\begin{equation}
        e^{i\phi}\ket{b}\otimes e^{i\phi'}\ket{b'}= e^{i\phi'}\ket{b}\otimes e^{i\phi}\ket{b'}= e^{i(\phi+\phi')}\ket{bb'} \quad . 
\end{equation}
When this state is mapped to the probability space, due to an additional redundancy in the mapping, the following simplex vectors are not equivalent to each other:
\begin{equation}
    \vec{P}_{b}(e^{i\phi})\otimes\vec{P}_{b'}(e^{i\phi'})\neq\vec{P}_{b}(e^{i\phi'})\otimes\vec{P}_{b'}(e^{i\phi})\neq\vec{P}_{b}(e^{i(\phi+\phi')})\otimes\vec{p}_{b'}\neq\vec{p}_{b}\otimes\vec{P}_{b'}(e^{i(\phi+\phi')}) \quad .
\end{equation}
Each permutation of the phase in the simplex version can be defined as a different ordering of phases. We call each specific ordering of the phases  by $\sigma$. The operations on the simplex vectors can be defined so as to follow the transformation in the quantum case for each permutation in Eq.~(124). The only difference will be the way we store the absolute phase information of the quantum state in the simplex vectors. Each distinct ordering ($\sigma$, the specific permutation of phases) will define a different set of intermediate states of the simplex vectors under the same set of operations. The orderings that provide set of intermediate states that are equivalent to how quantum states transform are the ones in which we collect all the phases in a single overall phase. We refer to these specific orderings for the two qubit case as $\sigma=1$ and $\sigma=2$, respectively, defined as:
\begin{align}
    &\sigma=1:\quad \vec{P}^{(1)}_{bb'}(e^{i(\phi+\phi')})=\vec{P}_{b}(e^{i(\phi+\phi')})\otimes\vec{p}_{b'}\\
    \text{and, }&\sigma=2:\quad \vec{P}^{(2)}_{bb'}(e^{i(\phi+\phi')})=\vec{p}_{b}\otimes\vec{P}_{b'}(e^{i(\phi+\phi')})
\end{align}
Importantly, as we will prove below, the final measurement outcomes for the simplex states do not depend on which specific phase ordering that we choose. This is similar to the quantum case where the measurements of any observable for a given quantum state do not depend on the absolute phase of the wavefunction. We believe it is interesting that the absolute phase information of the quantum wavefunction can be stored in different phase orderings of the simplex vectors. We leave a detailed study of full implications of this for future work. For the subsequent sections we will only be concerned with the phase orderings $\sigma=1$ and $\sigma=2$ as described above where all the absolute phases are combined in a single overall phase factor. Moreover, in the following subsection we specify linear operations $\Gamma^{(n)}_{\sigma}$ and $\Omega^{(n)}_{\sigma}$ that would allow us to achieve and switch between these different ordered states.

We now introduce affine transformations that allow us to switch between different phase-ordered states.  Let $\vec{P}$ be the state where the absolute phase information of each qubit is stored in the corresponding simplex vector, i.e.,
\begin{equation}
    \vec{P}=\vec{P}_{b}(e^{i\phi})\otimes\vec{P}_{b'}(e^{i\phi'})
\end{equation}
Using the definitions and notation introduced in Section~III, Eq~(7) and Eq~(8), this state can be rewritten in the following form:
\begin{equation}
    \vec{P}=(\cos\phi\,\vec{\gamma}+\sin\phi\,\vec{\gamma}')\otimes\ket{b}\otimes(\cos\phi'\,\vec{\gamma}+\sin\phi'\,\vec{\gamma}')\otimes\ket{b'}
\end{equation}
if $\tilde{\omega}$ is a permutation operation with the following action, 
\begin{equation}
\tilde{\omega}\cdot(\vec{u}_{2\times1}\otimes\vec{v}_{4\times1})=\vec{v}_{4\times1}\otimes\vec{u}_{2\times1}
\end{equation}
then $\vec{P}$ can be rewritten with the help of $\tilde{\omega}$ as,
\begin{align}
    \vec{P}&=(I_{4\times4}\otimes\overbrace{\tilde{\omega}^{\mathsf{T}}\tilde{\omega}}\otimes I_{2\times2})\cdot\left[(\cos\phi\,\vec{\gamma}+\sin\phi\,\vec{\gamma}')\otimes\overbrace{\ket{b}\otimes(\cos\phi'\,\vec{\gamma}+\sin\phi'\,\vec{\gamma}')}\otimes\ket{b'}\right]\\
    &=(I_{4\times4}\otimes\tilde{\omega}^{\mathsf{T}}\otimes I_{2\times2})\cdot\left[(\cos\phi\,\vec{\gamma}+\sin\phi\,\vec{\gamma}')\otimes(\cos\phi'\,\vec{\gamma}+\sin\phi'\,\vec{\gamma}')\otimes\ket{bb'}\right]
\end{align}
which is equivalent to:
\begin{equation}
    \vec{P}=(I_{4\times4}\otimes\tilde{\omega}^{\mathsf{T}}\otimes I_{2\times2})\cdot\left(\begin{array}{r}
			\vec{\xi}\\
			-\vec{\xi}\\
			\vec{\zeta}\\
			-\vec{\zeta}
		\end{array}\right)\otimes\ket{bb'}
\end{equation}
where, 
\begin{equation}
    \vec{\xi}=\left(\begin{array}{r}
		\cos\phi\,\cos\phi'\\
		-\cos\phi\,\cos\phi'\\
		\cos\phi\,\sin\phi'\\
		-\cos\phi\,\sin\phi'
	\end{array}\right),\quad \vec{\zeta}=\left(\begin{array}{r}
		\sin\phi\,\cos\phi'\\
		-\sin\phi\,\cos\phi'\\
		\sin\phi\,\sin\phi'\\
		-\sin\phi\,\sin\phi'
	\end{array}\right),
\end{equation}
and, 
\begin{equation}
    \tilde{\omega}=\left(\begin{array}{cccccccc}
		1&0&0&0&0&0&0&0\\
        0&0&1&0&0&0&0&0\\
        0&0&0&0&1&0&0&0\\
        0&0&0&0&0&0&1&0\\
        0&1&0&0&0&0&0&0\\
        0&0&0&1&0&0&0&0\\
        0&0&0&0&0&1&0&0\\
        0&0&0&0&0&0&0&1
	\end{array}\right),\quad \tilde{\omega}^{\mathsf{T}}=\tilde{\omega}^2=\left(\begin{array}{cccccccc}
		1&0&0&0&0&0&0&0\\
        0&0&0&0&1&0&0&0\\
        0&1&0&0&0&0&0&0\\
        0&0&0&0&0&1&0&0\\
        0&0&1&0&0&0&0&0\\
        0&0&0&0&0&0&1&0\\
        0&0&0&1&0&1&0&0\\
        0&0&0&0&0&0&0&1
	\end{array}\right)
\end{equation}
On the application of the phase ordering operations the resultant states should be of the form $\vec{P}^{(1)}=\vec{P}_{b}(e^{i(\phi+\phi')})\otimes\vec{p}_{b'}$,  
\begin{equation}
    \vec{P}^{(1)}=\left(\begin{array}{r}
			\cos(\phi+\phi')\\
			-\cos(\phi+\phi')\\
			\sin(\phi+\phi')\\
			-\sin(\phi+\phi')
		\end{array}\right)\otimes\ket{b}\otimes\vec{\gamma}\otimes\ket{b'}=(I_{4\times4}\otimes\tilde{\omega}^{\mathsf{T}}\otimes I_{2\times2})\cdot\left(\begin{array}{r}
			\cos(\phi+\phi')\vec{\gamma}\\
			-\cos(\phi+\phi')\vec{\gamma}\\
			\sin(\phi+\phi')\vec{\gamma}\\
			-\sin(\phi+\phi')\vec{\gamma}
		\end{array}\right)\otimes\ket{bb'}
\end{equation}
and $\vec{P}^{(2)}=\vec{p}_{b}\otimes\vec{P}_{b'}(e^{i(\phi+\phi')})$, 
\begin{equation}
    \vec{P}^{(2)}=\vec{\gamma}\otimes\ket{b}\otimes\left(\begin{array}{r}
			\cos(\phi+\phi')\\
			-\cos(\phi+\phi')\\
			\sin(\phi+\phi')\\
			-\sin(\phi+\phi')
		\end{array}\right)\otimes\ket{b'}=(I_{4\times4}\otimes\tilde{\omega}^{\mathsf{T}}\otimes I_{2\times2})\cdot(\iota\otimes I_{4\times4})\cdot\left(\begin{array}{r}
			\cos(\phi+\phi')\vec{\gamma}\\
			-\cos(\phi+\phi')\vec{\gamma}\\
			\sin(\phi+\phi')\vec{\gamma}\\
			-\sin(\phi+\phi')\vec{\gamma}
		\end{array}\right)\otimes\ket{bb'}
\end{equation}
where, $\tilde{\iota}$ is a symmetric permutation matrix and has action like $\tilde{\omega}$, 
\begin{align}
    &\tilde{\iota}\cdot(u_{4\times1}\otimes v_{4\times1})=v_{4\times1}\otimes u_{4\times1},\\
    &\tilde{\iota}=\tilde{\iota}^{\mathsf{T}}=\left(
\begin{array}{cccccccccccccccc}
 1 & 0 & 0 & 0 & 0 & 0 & 0 & 0 & 0 & 0 & 0 & 0 & 0 & 0 & 0 & 0 \\
 0 & 0 & 0 & 0 & 1 & 0 & 0 & 0 & 0 & 0 & 0 & 0 & 0 & 0 & 0 & 0 \\
 0 & 0 & 0 & 0 & 0 & 0 & 0 & 0 & 1 & 0 & 0 & 0 & 0 & 0 & 0 & 0 \\
 0 & 0 & 0 & 0 & 0 & 0 & 0 & 0 & 0 & 0 & 0 & 0 & 1 & 0 & 0 & 0 \\
 0 & 1 & 0 & 0 & 0 & 0 & 0 & 0 & 0 & 0 & 0 & 0 & 0 & 0 & 0 & 0 \\
 0 & 0 & 0 & 0 & 0 & 1 & 0 & 0 & 0 & 0 & 0 & 0 & 0 & 0 & 0 & 0 \\
 0 & 0 & 0 & 0 & 0 & 0 & 0 & 0 & 0 & 1 & 0 & 0 & 0 & 0 & 0 & 0 \\
 0 & 0 & 0 & 0 & 0 & 0 & 0 & 0 & 0 & 0 & 0 & 0 & 0 & 1 & 0 & 0 \\
 0 & 0 & 1 & 0 & 0 & 0 & 0 & 0 & 0 & 0 & 0 & 0 & 0 & 0 & 0 & 0 \\
 0 & 0 & 0 & 0 & 0 & 0 & 1 & 0 & 0 & 0 & 0 & 0 & 0 & 0 & 0 & 0 \\
 0 & 0 & 0 & 0 & 0 & 0 & 0 & 0 & 0 & 0 & 1 & 0 & 0 & 0 & 0 & 0 \\
 0 & 0 & 0 & 0 & 0 & 0 & 0 & 0 & 0 & 0 & 0 & 0 & 0 & 0 & 1 & 0 \\
 0 & 0 & 0 & 1 & 0 & 0 & 0 & 0 & 0 & 0 & 0 & 0 & 0 & 0 & 0 & 0 \\
 0 & 0 & 0 & 0 & 0 & 0 & 0 & 1 & 0 & 0 & 0 & 0 & 0 & 0 & 0 & 0 \\
 0 & 0 & 0 & 0 & 0 & 0 & 0 & 0 & 0 & 0 & 0 & 1 & 0 & 0 & 0 & 0 \\
 0 & 0 & 0 & 0 & 0 & 0 & 0 & 0 & 0 & 0 & 0 & 0 & 0 & 0 & 0 & 1 \\
\end{array}
\right)
\end{align}
\begin{figure}[b]
    \centering
    \includegraphics[width=0.5\textwidth]{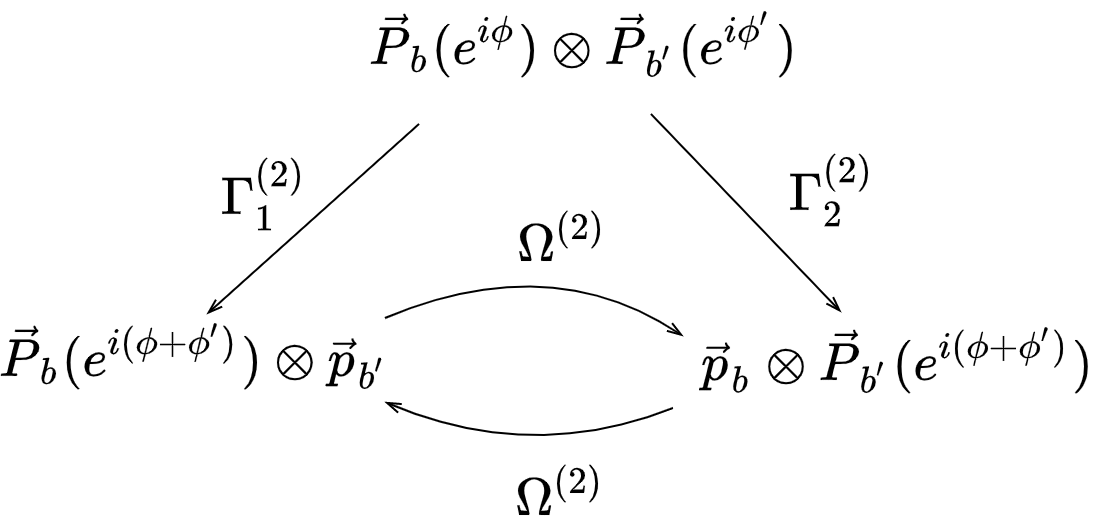}
    \caption{Phase ordering operations $\Gamma^{(2)}_{\sigma}$ and the order switching operation $\Omega^{(2)}$.}
\end{figure}
This implies that to be able to switch between different phase-orderings, we need to form linear combinations of the components of $\vec{\xi}$ and $\vec{\zeta}$ in $\vec{P}$. This can be done by various linear operations, we specify the following for the phase ordering operations, 
\begin{equation*}
    \tilde{\Gamma}=\left( \begin{array}{cccc|cccc|cccc|cccc}
I_{4\times4} & O & O & O &  O & O & O & O &  O & O & O & I_{4\times4} &  O & O & O & O \\
O & I_{4\times4} & O & O &  O & O & O & O &  O & O & I_{4\times4} & O &  O & O & O & O \\
O & O & I_{4\times4} & I_{4\times4} &  O & O & O & O &  O & O & O & O &  O & O & O & O \\
O & O & I_{4\times4} & I_{4\times4} &  O & O & O & O &  O & O & O & O &  O & O & O & O \\
\hline
O & O & O & O &  I_{4\times4} & O & O & O &  O & O & O & O &  O & O & O & I_{4\times4} \\
O & O & O & O &  O & I_{4\times4} & O & O &  O & O & O & O &  O & O & I_{4\times4} & O \\
O & O & O & O &  O & O & I_{4\times4} & I_{4\times4} &  O & O & O & O &  O & O & O & O \\
O & O & O & O &  O & O & I_{4\times4} & I_{4\times4} &  O & O & O & O &  O & O & O & O \\
\hline
O & O & I_{4\times4} & O &  O & O & O & O &  I_{4\times4} & O & O & O &  O & O & O & O \\
O & O & O & I_{4\times4} &  O & O & O & O &  O & I_{4\times4} & O & O &  O & O & O & O \\
O & O & O & O &  O & O & O & O &  O & O & I_{4\times4} & I_{4\times4} &  O & O & O & O \\
O & O & O & O &  O & O & O & O &  O & O & I_{4\times4} & I_{4\times4} &  O & O & O & O \\
\hline
O & O & O & O &  O & O & I_{4\times4} & O &  O & O & O & O &  I_{4\times4} & O & O & O \\
O & O & O & O &  O & O & O & I_{4\times4} &  O & O & O & O &  O & I_{4\times4} & O & O \\
O & O & O & O &  O & O & O & O &  O & O & O & O &  O & O & I_{4\times4} & I_{4\times4} \\
O & O & O & O &  O & O & O & O &  O & O & O & O &  O & O & I_{4\times4} & I_{4\times4} 
\end{array} \right)
\end{equation*}
using this and transforming back we obtain the required ordering operations,
\begin{align}
    &\Gamma^{(2)}_1=(I_{4\times4}\otimes\tilde{\omega}^{\mathsf{T}}\otimes I_{2\times2})\cdot\tilde{\Gamma}\cdot(I_{4\times4}\otimes\tilde{\omega}\otimes I_{2\times2}),\\
    &\Gamma^{(2)}_2=(I_{4\times4}\otimes\tilde{\omega}^{\mathsf{T}}\otimes I_{2\times2})\cdot(\iota\otimes I_{4\times4})\cdot\tilde{\Gamma}\cdot(\iota\otimes I_{4\times4})\cdot(I_{4\times4}\otimes\tilde{\omega}\otimes I_{2\times2})\quad .
\end{align} 
We note that by using the above relations, we can transform between $\Gamma^{(2)}_{1}$ and $\Gamma^{(2)}_{2}$ using a unique symmetric orthogonal transformation matrix, $\Omega^{(2)}$, which we refer to as an order switching operator for two $\vec{p}$ vectors:
\begin{align}
    &\Omega^{(2)}=(I_{4\times4}\otimes\tilde{\omega}^{\mathsf{T}}\otimes I_{2\times2})\cdot(\tilde{\iota}\otimes I_{4\times4})\cdot(I_{4\times4}\otimes\tilde{\omega}\otimes I_{2\times2}),\\
    &\Gamma^{(2)}_{2}=\Omega^{(2)}\cdot\Gamma^{(2)}_{1}\cdot\Omega^{(2)}, \quad \Gamma^{(2)}_{1}=\Omega^{(2)}\cdot \Gamma^{(2)}_{2}\cdot \Omega^{(2)} \quad . 
\end{align}
The diagram  that is shown in FIG.~9 summarizes the above operations for transforming between different phase-orderings when two qubits are mapped to the corresponding simplex vectors. Finally, the affine transformations corresponding to these operations are denoted by $\tilde{\Gamma}^{(2)}_{\sigma}$ and $\tilde{\Omega}^{(2)}$: 
\begin{equation}
    \tilde{\Gamma}^{(n)}_{\sigma}=T[\Gamma^{(n)}_{\sigma}],\quad 
    \tilde{\Omega}^{(n)}_{\sigma}=T[\Omega^{(n)}_{\sigma}]\quad .
\end{equation}
Above we have discussed transformation matrices to switch between different phase-ordered vectors, when any two qubits are mapped to the probability simplex. When the number of qubits is larger than two, the required order generating and switching operations can be obtained from the ones that are used in the two qubit case. As an example let's consider the case of four qubits mapped to the probability space. The first order operation in this case is given by the following expression, 
\begin{align}
    \Gamma^{(4)}_{1}=(\Gamma^{(2)}_1\otimes I_{8\times8}\otimes I_{8\times8})\cdot(I_{8\times8}\otimes \Gamma^{(2)}_1\otimes I_{8\times8})\cdot(I_{8\times8}\otimes I_{8\times8}\otimes\Gamma^{(2)}_1)
\end{align}
The circuit diagram for producing this operation is shown in FIG.~10. Now that we have the first order operation we can find all the other ones using three order switching operations $\Omega^{(4)}_{1},\Omega^{(4)}_2$ and $\Omega^{(4)}_3$. The commutative diagram in FIG.~11 provides a visual for the action of the three order switching operations.
\begin{figure}
    \centering
    \includegraphics[width=0.6\textwidth]{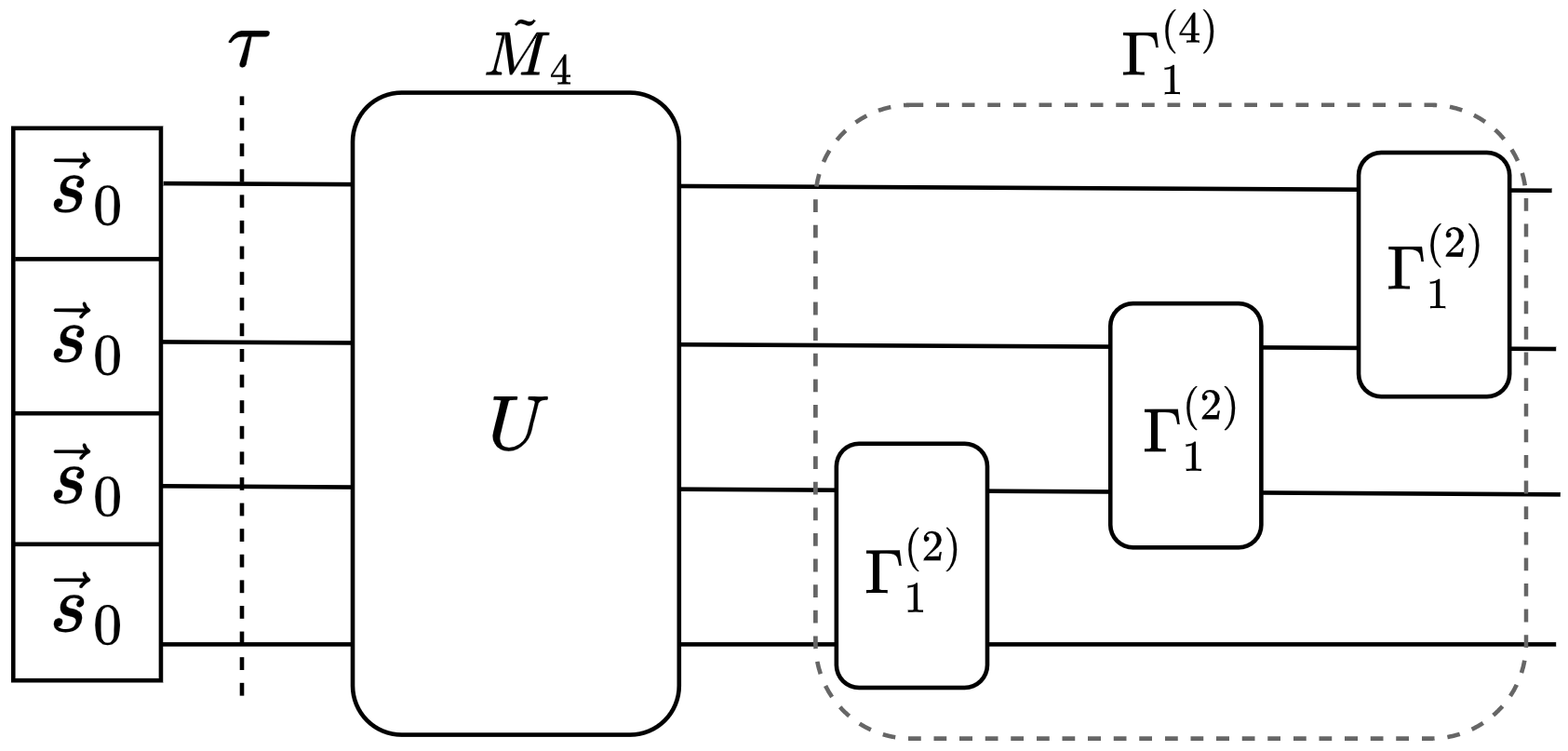}
    \caption{A circuit diagram for the application of first phase ordering operation in the case of four qubits.}
\end{figure}
\begin{figure}
    \centering
    \includegraphics[width=0.9\textwidth]{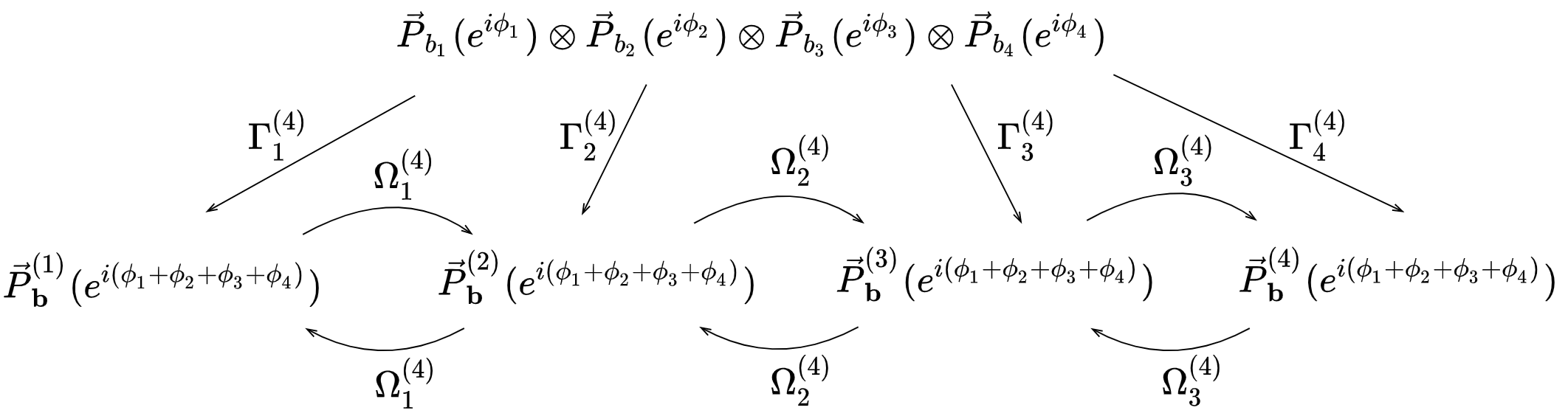}
    \caption{Action of the three order switching operations in the case of four qubit simplex states.}
\end{figure}
Therefore, each of the switching operation can be decomposed as, 
\begin{align}
    \Omega^{(4)}_{1}=\Omega^{(2)}\otimes I_{8\times8}\otimes I_{8\times8}\\
    \Omega^{(4)}_{2}=I_{8\times8}\otimes\Omega^{(2)}\otimes I_{8\times8}\\
    \Omega^{(4)}_{3}=I_{8\times8}\otimes I_{8\times8}\otimes\Omega^{(2)}
\end{align}
using these we can then obtain all the other phase ordering operations $\Gamma^{(4)}_{\sigma},\sigma\ge 2$, 
\begin{align}
    &\Gamma^{(4)}_{2}=\Omega^{(4)}_{1}\cdot\Gamma^{(4)}_{1}\cdot\Omega^{(4)}_{1}\\
    &\Gamma^{(4)}_{3}=\Omega^{(4)}_{2}\cdot\Gamma^{(4)}_{2}\cdot\Omega^{(4)}_{2}\\
    &\Gamma^{(4)}_{4}=\Omega^{(4)}_{3}\cdot\Gamma^{(4)}_{3}\cdot\Omega^{(4)}_{3}\quad .
\end{align}
Finally, we can comment on the mapping of a general $n$-qubit quantum state. Because we have the ability to put a state in a definite phase ordering, we additionally define phase ordered maps $\varphi^{\sigma}_n$ that have action as described below. Given some general $n$-qubit quantum state in the logical basis:
\begin{equation}
    \ket{\psi}=\sum_{{\bf{q}}\in\mathbb{B}^n}c_{\bf{q}}\ket{\bf{q}}
\end{equation}
and $\sigma\in\mathbb{Z}_n+1=\{m|\,1\leqslant m\leqslant n,m\in\mathbb{Z}^+\}$, 
\begin{equation}
    \varphi^{\sigma}_{n}\ket{\psi}=\vec{s}^{(\sigma)}(\psi)\doteq\frac{1}{8^{n}}(\vec{u}^{\otimes_n}+\sum_{{\bf{q}}\in\mathbb{B}^n}\vec{P}^{(\sigma)}_{\bf{q}}(c_{\bf{q}}))
\end{equation}

\section{Appendix E. Measurement invariance of phase ordering}
In this Appendix we show that the measurement of an observable in the logical basis is independent of the phase ordering chosen for the states. Moreover, we also prove that this allows us to measure states in any choice of basis other than the logical basis.

From the main text we know that for a single-qubit quantum state $\ket{\psi}=\sum_{u\in\mathbb{B}}c_u\ket{u}$, its mapped simplex state: 
\begin{equation}
    \vec{s}=\varphi\ket{\psi}=\frac{1}{8}\left(\vec{u}+\overbrace{\sum_{u\in\mathbb{B}}\vec{P}_{u}(c_u)}^{\vec{P}}\right) \quad ,
\end{equation} 
and a quantum observable $\hat{A}$, the following connection holds (cf. Eq.~(30)): 
\begin{equation}
    \bra{\psi}\hat{A}\ket{\psi}=\frac{1}{2}(\vec{P}^{\mathsf{T}}\cdot\tilde{M}[\hat{A}]\cdot\vec{P}) \quad ,
\end{equation}

We know that the set of Pauli operators and Identity matrix $\mathcal{P}=\{\hat{\sigma}_0=\hat{I}_{2\times2},\hat{\sigma}_1=\hat{X},\hat{\sigma}_2=\hat{Y},\hat{\sigma}_3=\hat{Z}\}$ forms a complete and orthogonal basis for any $2\times2$ operator. Hence, for any $2\times2$ quantum observable $\hat{A}$ we would have the following decomposition, 
\begin{equation}
    \hat{A}=\sum_{i=0}^{3}a^{i}\,\hat{\sigma}_{i} \quad ,
\end{equation}
where, each of the coefficients $a^i$ are real (because $\hat{A}$ is Hermitian) and can be obtained by the trace formula: $a^i=\Tr(\hat{\sigma}_{i}\hat{A})$. Using this fact and expanding the state $\ket{\psi}$ and $\vec{s}$ on the left and right hand side of Eq.~(154) we obtain: 
\begin{equation}
\sum_{u,v\in\mathbb{B}^2}\sum_{i=0}^{3}c_u^{*}c_v\,a^{i}\bra{u}\hat{\sigma}_{i}\ket{v}=\sum_{u,v\in\mathbb{B}^2}\sum_{i=0}^{3}c_u^{*}\,a^{i}\sigma^{uv}_{i}\,c_v=\frac{1}{2}\left(\sum_{u,v\in\mathbb{B}^2}\vec{P}^{\mathsf{T}}_{u}(c_u)\cdot\tilde{M}[\sum_{i=0}^{3}a^{i}\hat{\sigma}_i]\cdot\vec{P}_{v}(c_v)\right) \quad ,
\end{equation}
from the definition of $\tilde{M}$ operator map we see that:
\begin{equation}
    \tilde{M}[\sum_{i=0}^{3}a^{i}\hat{\sigma}_i]=\sum_{i=0}^{3}a^{i}\tilde{M}[\hat{\sigma}_i],\;\because\; a_i\in\mathbb{R},\,\forall\,i\in\{0,1,2,3\}
\end{equation}
as a consequence we can subsequently write the following equivalence for any $(u,v)\in\mathbb{B}^2$, $(c_u,c_v)\in \mathbb{C}^2$ and $i\in\{0,1,2,3\}$: 
\begin{equation}
    2\,c^{*}_{u}c_v\sigma^{uv}_{i}=\vec{P}^{\mathsf{T}}_{u}(c_u)\cdot\tilde{M}[\hat{\sigma}_i]\cdot\vec{P}_v(c_v) \quad . 
\end{equation}

With this notation and considerations for the single-qubit case in mind, let's move on to the $n$-qubit case. The quantum state $\ket{\psi}=\sum_{{\bf{q}}\in\mathbb{B}^n}c_{\bf{q}}\ket{\bf{q}}$ under some phase ordering $\omega\in\{1,2,\dots,n\}$ can be mapped to a simplex state as follows: 
\begin{equation}
    \varphi^{\omega}_{n}\ket{\psi}=\vec{s}^{(\omega)}=\frac{1}{8^{n}}\left(\vec{u}^{\otimes_{n}}+\overbrace{\sum_{{\bf{q}}\in\mathbb{B}^n}\vec{P}^{(\omega)}_{\bf{q}}(c_{\bf{q}})}^{\vec{P}^{(\omega)}}\right) \quad .
\end{equation}
Now for a given $2^{n}\times2^{n}$ quantum observable $\hat{A}$ which we can decompose as:
\begin{equation}
\hat{A}=\sum_{\mu,\dots,\zeta,\dots,\eta}a^{\mu,\dots,\zeta,\dots,\eta}\,\overbrace{\hat{\sigma}_{\mu}\otimes\cdots\otimes\hat{\sigma}_{\zeta}}^{\omega\text{ terms}}\otimes\cdots\otimes\hat{\sigma}_{\eta} \quad ,
\end{equation}
where $a^{\mu,\dots,\zeta,\dots,\eta}\in\mathbb{R}, \forall (\mu,\dots,\zeta,\dots,\eta)\in\{0,1,2,3\}^{n}$, we follow through the below steps: 
\begin{align*}
    &\frac{1}{2^{n}}((\vec{P}^{(\omega)})^{\mathsf{T}}\cdot\tilde{M}_n[\hat{A}]\cdot\vec{P}^{(\omega)})=\frac{1}{2^{n}}\left(\sum_{{\bf{q}},{\bf{k}}}(\vec{P}^{(\omega)}_{\bf{q}}(c_{\bf{q}}))^{\mathsf{T}}\cdot\tilde{M}_n[\hat{A}]\cdot\vec{P}^{(\omega)}_{\bf{k}}(c_{\bf{k}})\right)\\
    &=\frac{1}{2^{n}}\left(\sum_{{\bf{q}},{\bf{k}}}(\vec{P}^{(\omega)}_{\bf{q}}(c_{\bf{q}}))^{\mathsf{T}}\cdot\tilde{M}_n\left[\sum_{\mu,\dots,\zeta,\dots,\eta}a^{\mu,\dots,\zeta,\dots,\eta}\hat{\sigma}_{\mu}\otimes\cdots\otimes\hat{\sigma}_{\zeta}\otimes\cdots\otimes\hat{\sigma}_{\eta}\right]\cdot\vec{P}^{(\omega)}_{\bf{k}}(c_{\bf{k}})\right)\\
    &=\frac{1}{2^{n}}\left(\sum_{{\bf{q}},{\bf{k}}}\sum_{\mu,\dots,\zeta,\dots,\eta}a^{\mu,\dots,\zeta,\dots,\eta}\, (\vec{p}^{\mathsf{T}}_{q_1}\otimes\cdots\otimes\vec{P}^{\mathsf{T}}_{q_\omega}(c_{\bf{q}})\otimes\cdots\otimes\vec{p}^{\mathsf{T}}_{q_n})\cdot(\tilde{M}[\hat{\sigma}_{\mu}]\otimes\cdots\otimes\tilde{M}[\hat{\sigma}_{\zeta}]\otimes\cdots\otimes\tilde{M}[\hat{\sigma}_{\eta}])\cdot\right.\\
    &\quad \quad \quad \quad (\vec{p}_{k_1}\otimes\cdots\otimes\vec{P}_{k_\omega}(c_{\bf{k}})\otimes\cdots\otimes\vec{p}_{k_n})\bigg)\quad (\because\; a^{\mu,\dots,\zeta,\dots,\eta}\in\mathbb{R}, \forall (\mu,\dots,\zeta,\dots,\eta)\in\{0,1,2,3\}^{n})\\
    &=\frac{1}{2^{n}}\left(\sum_{{\bf{q}},{\bf{k}}}\sum_{\mu,\dots,\zeta,\dots,\eta}a^{\mu,\dots,\zeta,\dots,\eta}\; (\vec{p}^{\mathsf{T}}_{q_1}\cdot\tilde{M}[\hat{\sigma}_{\mu}]\cdot\vec{p}_{k_1})\;\cdots\;(\vec{P}^{\mathsf{T}}_{q_\omega}(c_{\bf{q}})\cdot\tilde{M}[\hat{\sigma}_{\zeta}]\cdot\vec{P}_{k_\omega}(c_{\bf{k}}))\;\cdots\;(\vec{p}^{\mathsf{T}}_{q_n}\cdot\tilde{M}[\hat{\sigma}_{\eta}]\cdot\vec{p}_{k_n})\right)\\
    &=\frac{1}{2^{n}}\left(\sum_{{\bf{q}},{\bf{k}}}\sum_{\mu,\dots,\zeta,\dots,\eta}a^{\mu,\dots,\zeta,\dots,\eta}\; (2\,\sigma^{q_1k_1}_{\mu})\;\cdots\;(2\,c^{*}_{\bf{q}}\,\sigma^{q_{\omega}k_{\omega}}_{\zeta}\,c_{\bf{k}})\;\cdots\;(2\,\sigma^{q_nk_n}_{\eta})\right)\quad (\text{cf. Eq.~(158)})\\
    &=\left(\sum_{{\bf{q}},{\bf{k}}}c^{*}_{\bf{q}}\bra{\bf{q}}\left(\sum_{\mu,\dots,\zeta,\dots,\eta}a^{\mu,\dots,\zeta,\dots,\eta}\; \hat{\sigma}_{\mu}\otimes\;\cdots\;\otimes\hat{\sigma}_{\zeta}\otimes\;\cdots\;\otimes\hat{\sigma}_{\eta}\right)c_{\bf{k}}\ket{\bf{k}}\right)=\bra{\psi}\hat{A}\ket{\psi}=\langle\hat{A}\rangle
\end{align*}
Therefore, we observe that regardless of the chosen phase order $\omega$ we obtain the correct quantum measurement value $\langle\hat{A}\rangle$ proving the validity of the premise of this appendix.

We finally note that, in place of the logical basis states let us suppose that we have the basis set: $\{\ket{\phi_i}|\,i\in\{0,1,\dots,2^n-1\}\}$, and the operator $\hat{\Phi}$ that links the logical basis to this new basis of states:
\begin{equation}
    \hat{\Phi}\ket{\bf{q}}=\ket{\phi_q}, q=[{\bf{q}}]\in\{0,1,\dots,2^n-1\}\quad .
\end{equation}
We can then construct a quantum observable $\hat{A}_{q}=\hat{\Phi}\hat{M}_{\bf{q}}\hat{\Phi}^{\dagger}=\op{\phi_q}$ which measures any simplex state $\vec{s}=\varphi_n\ket{\psi}$ (note here that the phase order has not been specified because of its irrelevance) in the new chosen basis set $\{\ket{\phi_i}|\,i\in\{0,1,\dots,2^n-1\}\}$:
\begin{equation}
    \langle T[\hat{A}_{q}]\rangle_{\vec{s}}=\vec{s}\cdot\,T[\hat{A}_{q}](\vec{s})=\frac{1}{8^n}(1+\frac{1}{4^n}|\braket{\phi_q}{\psi}|^2),\quad q=[{\bf{q}}]\in\{0,1,\dots,2^{n}-1\}
\end{equation}


\newpage 
\begin{references}

\bibitem{nielsen} M. A. Nielsen and I. L. Chuang, {\it Quantum Computation and Quantum Information} (Cambridge University
Press, 2000).

\bibitem{divincenzo1} D. P. DiVincenzo, {\it Quantum Computation}, Science {\bf 270}, 255 (1995).

\bibitem{divincenzo2} C. H. Bennett and D. P. DiVincenzo, {\it Quantum Information and Computation}, Nature {\bf 404}, 247 (2000).

\bibitem{vazirani} E. Bernstein and U. Vazirani, {\it Quantum Complexity Theory}, SIAM J. Comput. {\bf 26}, 1411 (1997).

\bibitem{shor} P. W. Shor, {\it Polynomial-time Algorithms for Prime Factorization and Discrete Logarithms on a Quantum
Computer}, SIAM J. Comp. {\bf 26}, 1484 (1997).

\bibitem{ekert} A. Ekert and R. Jozsa, {\it Quantum Computation and Shor’s Factoring Algorithm}, Rev. Mod. Phys. {\bf 68},
733 (1996).

\bibitem{coppersmith} D. Coppersmith, {\it An Approximate Fourier Transform Useful in Quantum Factoring}, Technical Report RC19642, IBM, arXiv:quant-ph/0201067 (1994). 

\bibitem{deutsch} D. Deutsch and R. Jozsa, {\it Rapid Solution of Problems by Quantum Computation}, Proc. R. Soc. London A {\bf 439}, 553 (1992). 

\bibitem{grover} L. K. Grover, {\it Quantum Mechanics Helps in Searching for a Needle in Haystack}, Phys. Rev. Lett. {\bf 79}, 325 (1997).

\bibitem{abrams} D. S. Abrams and S. Lloyd, {\it Quantum Algorithm Providing Exponential Speed Increase for Finding Eigenvalues and Eigenvectors}, Phys. Rev. Lett. {\bf 83}, 5162 (1999).

\bibitem{watrous} J. Watrous, {\it Quantum Computational Complexity}, arXiv:0804.3401 [quant-ph] (2008). 

\bibitem{wetterich1} C. Wetterich, {\it Quantum Computing with Classical Bits}, Nuclear Physics B {\bf 948}, 114776 (2019). 

\bibitem{wetterich2} C. Wetterich, {\it Quantum Formalism for Classical Statistics}, Annals of Physics {\bf 393}, 1 (2018). 

\bibitem{duan} X. Gao and L. Duan, {\it Efficient Classical Simulation of Noisy Quantum Computation}, arXiv:1810.03176v1 [quant:ph] (2018).

\bibitem{waintal} Y. Zhou, E. M. Stoudenmire, and X. Waintal, {\it What Limits the Simulation of Quantum Computers}, Phys. Rev. X {\bf 10}, 041038 (2020). 

\bibitem{jozsa} R. Jozsa and N. Linden, {\it On the Role of Entanglement in Quantum-Computational Speed-Up}, Proceedings: Mathematical, Physical and Engineering Sciences {\bf 459}, 2011 (2003). 

\bibitem{nest} M. van den Nest, {\it Universal Quantum Computation with Little Entanglement}, Phys. Rev. Lett. {\bf 110}, 060504 (2013). 

\bibitem{bauer} Bela Bauer, D. Wecker, Andrew J. Millis, Matthew B. Hastings, and Matthias Troyer, {\it Hybrid Quantum-Classical Approach to Correlated Materials}, Phys. Rev. X {\bf 6}, 031045 (2016).

\bibitem{monroe} D. Zhu, N. M. Linke, M. Benedetti, K. A. Landsman, N. H. Nguyen, C. H. Alderete1, A. Perdomo-Ortiz, N. Korda, A. Garfoot, C. Brecque, L. Egan, O. Perdomo, and  C. Monroe, {\it Training of Quantum Circuits on a Hybrid
Quantum Computer}, Sci. Adv. {\bf 5}, DOI: 10.1126/sciadv.aaw9918 (2019). 

\bibitem{preskillreview} J. Preskill, {\it Quantum computing in the NISQ era and beyond},  Quantum {\bf 2}, 79 (2018).

\bibitem{deniz_rudhy} D. D. Yavuz and A. Yadav, {\it Mapping of Quantum Systems to the Probability Simplex}, arXiv:2301.06572v1 [quant-ph] (2023). 

\bibitem{datta1} K. Y. Camsari, R. Faria, B. M. Sutton, and S. Datta, {\it Stochastic p-Bits for Invertible Logic}, Phys. Rev. X {\bf 7}, 031014 (2017).

\bibitem{datta2} K. Y. Camsari, B. M. Sutton, and S. Datta, {\it p-Bits for Probabilistic Spin Logic}, arXiv:1809:04028 [cs.ET] (2019).

\bibitem{fuchs1} C. A. Fuchs and R. Shack, {\it Quantum-Bayesian Coherence}, Rev. Mod. Phys. {\bf 85}, 1693 (2013). 

\bibitem{hardy} L. Hardy, {\it Quantum Theory From Five Reasonable Axioms}, arXiv:0101012 [quant-ph] (2001). 

\bibitem{barrett1} J. Barrett, {\it Information Processing in Generalized Probabilistic Theories}, Phys. Rev. A {\bf 75}, 032304 (2007). 

\bibitem{fuchs2} C. M. Caves, C. A. Fuchs, and R. Shack, {\it Quantum Probabilities as Bayesian Probabilities}, Phys. Rev. A {\bf 65}, 022305 (2002). 

\bibitem{fuchs3} J. B. DeBrota, C. A. Fuchs,1,2 and B. C. Stacey, {\it Symmetric informationally Complete Measurements Identify the Irreducible Difference Between Classical and Quantum Systems}, Phys. Rev. Research, {\bf 2}, 013074 (2020). 

\bibitem{boyd} N. Bent, H. Qassim, A. A. Tahir, D. Sych, G. Leuchs, L. L. Sánchez-Soto, E. Karimi, and R.W. Boyd, {\it Experimental Realization of Quantum Tomography of Photonic Qudits via Symmetric
Informationally Complete Positive Operator-Valued Measures}, Phys. Rev. X {\bf 5}, 041006 (2015). 

\bibitem{barnum} H. Barnum and A. Wilce, {\it Post-Classical Probability Theory}, arXiv:1205.3833 [quant-ph] (2013). 

\bibitem{masanes} L. Masanes and M. P. Muller, {\it A Derivation of Quantum Theory from Physical Requirements}, arXiv:1004.1483 [quant-ph] (2011). 

\bibitem{rau} J. Rau, {\it On Quantum vs. Classical Probability}, arXiv:0710.2119 [quant-ph] (2009). 


\bibitem{rudolph1} N. Harrigan and T. Rudolph, {\it Ontological models and the Interpretation of Contextuality}, arXiv:0709.4266 [quant-ph] (2007). 

\bibitem{spekkens1} R. W. Spekkens, {\it In Defense of the Epistemic View of Quantum States: a Toy Theory}, arXiv:0401052 [quant-h] (2005). 

\bibitem{bell2} J. S. Bell, {\it On the Problem of Hidden Variables in Quantum Mechanics}, Reviews of Modern Phys. {\bf  38}, 447 (1966).

\bibitem{bugajski} E. G. Beltrametti and S. Bugajski, {\it A Classical Extension of Quantum Mechanics}, J. Phys. A: Math. Gen. {\bf 8}, 3329 (1995).

\bibitem{kochen} S. Kochen and E. Specker, {\it The Problem of Hidden Variables in Quantum Mechanics}, Journal of Mathematics and Mechanics {\bf 17}, 59 (1967).

\bibitem{aaronson} S. Aaronson, {\it Quantum Computing and Hidden Variables},  Phys. Rev. A {\bf 71}, 032325 (2005). 

\bibitem{aerts} D. Aerts, {\it A Possible Explanation for the Probabilities of Quantum Mechanics}, J. Math. Phys. {\bf 27}, 202 (1985).

\bibitem{bartlett1} J. J. Wallman and S. D. Bartlett, {\it Non-negative Subtheories and Quasiprobability Representations of Qubits}, Phys. Rev. A {\bf 85}, 062121 (2012). 

\bibitem{bartlett2} H. Pashayan, J. J. Wallman, and S. D. Bartlett, {\it Estimating Outcome Probabilities of Quantum Circuits Using Quasiprobabilities}, Phys. Rev. Lett. {\bf 115}, 070501 (2015). 

\bibitem{bartlett3} H. Pashayan,  O. Reardon-Smith , K. Korzekwa,  and S. D. Bartlett, {\it Fast Estimation of Outcome Probabilities for Quantum Circuits}, PRX Quantum {\bf 3}, 020361 (2022). 

\bibitem{raussendorf1} R. Raussendorf, D. E. Browne, N. Delfosse, C. Okay, and J. Bermejo-Vega, {\it Contextuality and Wigner-Function Negativity in Qubit Quantum Computation}, Phys. Rev. A {\bf 95}, 052334 (2017). 

\bibitem{raussendorf2} N. Delfosse, P. A. Guerin, J. Bian, and R. Raussendorf, {\it Wigner Function Negativity and Contextuality in Quantum Computation on Rebits}, Phys. Rev. X {\bf 5}, 021003 (2015). 

\bibitem{eisert} A. Mari and J. Eisert, {\it Positive Wigner Functions Render Classical Simulation of Quantum Computation Efficient}, Phys. Rev. Lett. {\bf 109}, 230503 (2012). 

\bibitem{zhu} H. Zhu, {\it Quasiprobability Representations of Quantum Mechanics with Minimal Negativity}, Phys. Rev. Lett. {\bf 117}, 120404 (2016). 

\bibitem{wootters} K. S. Gibbons, M. J. Hoffman, and W. K. Wootters, {\it Discrete Phase Space Based on Finite Fields}, Phys. Rev. A {\bf 70}, 062101 (2004). 

\bibitem{definetti} B. de Finetti, {\it Theory of Probability} (Wiley, New York, 1990).

\bibitem{aharonov} D. Aharonov and M. Ben-Or, {\it Fault Tolerant Quantum Computation with Constant Error}, Proceedings of the 29th Annual ACM Symposium on the Theory of Computing, 176 (1997).

\bibitem{terhal} B. M. Terhal and G. Burkard, {\it Fault-Tolerant Quantum Computation for Local Non-Markovian Noise}, Phys. Rev. A {\bf 71}, 012336 (2005). 

\bibitem{aharonov2} D. Aharonov, A. Kitaev, and J. Preskill, {\it Fault-Tolerant Quantum Computation with Long-Range Correlated Noise}, Phys. Rev. Lett. {\bf 96}, 050504 (2006). 

\bibitem{preskill} H. K. Ng and J. Preskill, {\it Fault-Tolerant Quantum Computation Versus Gaussian Noise}, Phys. Rev. A {\bf 79}, 032318 (2009). 


\bibitem{mucciolo1} E. Novais and E. R. Mucciolo, {\it Surface Code Threshold in the Presence of Correlated Errors}, Phys. Rev. Lett. {\bf 110}, 010502 (2013). 

\bibitem{mucciolo2} E. Novais, A. J. Stanforth, and Eduardo R. Mucciolo, {\it Surface Code Fidelity at Finite Temperatures}, Phys. Rev. A {\bf 95}, 042339 (2017).

\bibitem{hutter} A. Hutter and D. Loss, {\it Breakdown of Surface-Code Error Correction Due to Coupling to a Bosonic Bath}, Phys. Rev. A {\bf 89}, 042334 (2014). 

\bibitem{lemberger} B. Lemberger and D. D. Yavuz, {\it Effect of Correlated Decay on Fault-Tolerant Quantum Computation}, Phys. Rev. A {\bf 96}, 062337 (2017). 

\bibitem{niklas} N. Johansson and J. Larsson, {\it Efficient classical simulation of the Deutsch - Jozsa and Simon’s algorithms}, Quantum Inf. Process. {\bf 16}, 233 (2017). 

\bibitem{mucciolo} C. Chamon and E. R. Mucciolo, {\it Virtual Parallel Computing and a Search Algorithm Using Matrix Product States}, Phys. Rev. Lett. {\bf 109}, 030503 (2012).  

\bibitem{pittenger} C. Cormick, E. F. Galvão, D. Gottesman, J. P. Paz, and A. O. Pittenger, {\it Classicality in Discrete Wigner Functions}, Phys. Rev. A {\bf 73}, 012301 (2006). 

\bibitem{zurek1} W. H. Zurek, Decoherence, Einselection, and the Quantum Origins of the Classical, Rev. Mod. Phys. {\bf 45}, 715 (2003). 

\bibitem{zurek2} W. H. Zurek, Quantum Darwinism, Nature Phys. {\bf 5}, 181 (2009). 

\bibitem{bell1} J. S. Bell, {\it Against Measurement}, Phys. World {\bf 3}, 33 (1990). 

\bibitem{realq1} M. O. Renou, D. Trillo, M. Weilenmann, T. P. Le, A. Tavakoli, N. Gisin, A. Acin, and  M. Navascues, {\it Quantum Theory Based on Real Numbers can be Experimentally Falsified}, Nature {\bf 600}, 625 (2021). 

\bibitem{realq2} Z. D. Li {\it al.}, {\it Testing Real Quantum Theory in an Optical Quantum Network}, Phys. Rev. Lett. {\bf 128}, 040402 (2022). 

\bibitem{barrett2} J. Barrett, N. Linden, S. Massar, S. Pironio, S. Popescu, and D. Roberts, {\it Nonlocal Correlations as an Information-Theoretic Resource}, Phys. Rev. A {\bf 71}, 022101 (2005). 

\bibitem{spekkens2} R. W. Spekkens, {\it Negativity and Contextuality are Equivalent Notions of Nonclassicality}, Phys. Rev. Lett. {\bf 101}, 020401 (2008). 

\bibitem{montina} A. Montina, {\it Exponential Complexity and Ontological Theories of Quantum Mechanics}, Phys. Rev. A {\bf 77}, 022104 (2008). 

\bibitem{rudolph2} P. G. Lewis, D. Jennings, J. Barrett, and T. Rudolph, {\it Distinct Quantum States Can Be Compatible with a Single State of Reality}, Phys. Rev. Lett. {\bf 109}, 150404 (2012). 

\bibitem{rudolph3} M. F. Pusey, J. Barrett, and T. Rudolph, {\it On the Reality of the Quantum State}, Nature Phys. {\bf 8}, 475 (2012). 

\end{references}
\end{document}